\documentclass{aa}  
\usepackage{graphicx}
\usepackage{txfonts}
 \usepackage{lscape}
 \usepackage{colortbl}
 \usepackage{natbib}
 \usepackage{verbatim} 
\usepackage{url}
 \bibpunct{(}{)}{;}{a}{}{,}
\begin{document}
   \title{A robust morphological classification of high-redshift galaxies using support vector machines on seeing limited images}
   \subtitle{II. Quantifying morphological k-correction in the COSMOS field at $1<z<2$: Ks band vs. I band
\thanks{Based on observations obtained at the Canada-France-Hawaii
    Telescope (CFHT) which is operated by the National Research
    Council of Canada, the Institut National des Sciences de l'Univers
    of the Centre National de la Recherche Scientifique of France, and
    the University of Hawaii.} 
}
   
   %\subtitle{I. Overviewing the $\kappa$-mechanism}
\author{M. Huertas-Company\inst{1,3},
          L. Tasca \inst{2},
          D. Rouan \inst{1}, 
          D. Pelat \inst{4},
         J.P. Kneib \inst{2},
          O. Le F\`evre \inst{2},
          P. Capak\inst{7},
          J. Kartaltepe\inst{8},
          A. Koekemoer\inst{10},
          H. J. McCracken\inst{5},
          M. Salvato\inst{6},
          D. B. Sanders\inst{8},
          C. Willott\inst{9}
           }

   \institute{LESIA-Paris Observatory, 
   5 Place Jules Janssen, 92195 Meudon, France\\
              \email{marc.huertas@obspm.fr}         
         \and
           LAM, CNRS-Universit\'e de Provence,
           38, rue Fr\'edŽric Joliot-Curie, 13388 Marseille cedex 13, France
           \and
           IAA-C/ Camino Bajo de Hu\'etor, 50 - 18008 Granada, Spain
           \and
           LUTH-Paris Observatory,
           5 Place Jules Janssen, 92195 Meudon, France
           \and
           IAP, CNRS-Universit\'e Pierre et Marie Curie,
           98, Boulevard Arago, F-75014 Paris, France
           \and
           Caltech, Pasadena, CA 91125,105-24 Caltech, Pasadena, CA 91125
           \and
           Spitzer Science Center, 314-6 Caltech, Pasadena, CA 91125,105-24 Caltech, Pasadena, CA 91125
           \and
           Institute for Astronomy, 2680 Woodlawn Drive, University of Hawaii, Honolulu, HI 96822
           \and
           Physics Department, University of Ottawa, 150 Louis Pasteur, MacDonald Hall, Ottawa, ON K1N 6N5, Canada
           \and
           Space Telescope Science Institute, 3700 San Martin Drive, Baltimore MD 21218, USA}

   \date{Received <date> / Accepted <date>}

% \abstract{}{}{}{}{} 
% 5 {} token are mandatory
 
  \abstract
   { Morphology is the most accessible tracer of galaxies physical
structure, but its interpretation in the framework of galaxy evolution
still remains problematic. Its quantification at high redshift requires deep high-angular resolution imaging,  which is why space data (HST) are usually employed. At $z>1$, the HST visible cameras however probe the UV flux, which is dominated by the emission of young stars, which could bias the estimated morphologies towards late-type systems.}
   {In this paper we quantify the effects of this \emph{morphological k-correction} at $1<z<2$ by comparing morphologies measured in the K and I-bands in the COSMOS area. The Ks-band data indeed have the advantage of probing old stellar populations in the rest frame for $z<2$, enabling determination of galaxy morphological types unaffected by recent star formation.}
 {  In paper I we presented a new non-parametric method of quantifying morphologies of galaxies on seeing-limited images based on support vector machines. Here we use this method to classify $\sim$$50\,000$ $Ks$ selected galaxies in the COSMOS area observed with WIRCam at CFHT.  We use a
10-dimensional volume, including 5 morphological parameters, 
and other characteristics of galaxies such as luminosity and redshift. The obtained classification is used to investigate the redshift distributions and number counts per morphological type up to $z\sim2$ and to compare them to the results obtained with HST/ACS in the I-band on the same objects. We associate to every galaxy with $Ks<21.5$ and $z<2$ a probability between 0 and 1 of being late-type or early-type. We use this value to assess the accuracy of our classification as a function of physical parameters of the galaxy and to correct for classification errors.  }
   {The classification is found to be reliable up to $z\sim2$. The mean probability is $p\sim0.8$. It decreases with redshift and with size, especially for the early-type population, but remains above $p\sim0.7$. The classification globally agrees with the one obtained using HST/ACS for $z<1$. Above $z\sim1$, the I-band classification tends to find less early-type galaxies than the Ks-band one by a factor $\sim$1.5,  which might be a consequence of morphological k-correction effects.}
   { We argue therefore that studies based on I-band HST/ACS classifications at $z>1$ could be underestimating the elliptical population. {Using our method in a $K_s \leq 21.5$ magnitude-limited
sample, we observe that the fraction of the early-type population is (21.9\% $\pm$ 8\%) at
$z\sim 1.5-2$ and (32.0\% $\pm$ 5\%) at the present time. We will discuss the
evolution of the fraction of galaxies in types from volume-limited
samples in a forthcoming paper.}}

   \keywords{galaxies: fundamental parameters -- galaxies: high redshift}

\authorrunning{Huertas-Company et al}
\titlerunning{NIR morphologies with WIRCam}
 \maketitle
%
%________________________________________________________________
 \nocite{Huertas-Company07, Cresci06, Baker03, McCracken00}
\section{Introduction}

In the local Universe, the distribution of galaxies is bimodal, primarily reflecting a relationship between color and morphology. On the one hand the spiral-like galaxies are gas-rich, form stars and are supported by the rotation of their stars and on the other hand the elliptical-like galaxies are gas-poor, do not form stars anymore and are supported by the velocity dispersion of the stars. This is the so-called elliptical-spiral Hubble sequence. A fundamental question in observational cosmology is how this bimodality appears throughout the history of the Universe. Classical approaches to tackle this question consist in studying the evolution of the luminosity (e.g.~\citealp{Il06}), the star-forming rate or the mass assembly (e.g.~\citealp{arn07, Bundy06}) for different morphological types. For that purpose, large samples of galaxies are required with a robust estimate of distances, luminosities and morphological types. 

However, the difficulty in quantifying morphology of high-redshift objects with a few simple, reliable measurements is still a major obstacle. The dependence on angular resolution and wavelength in fact turns the interpretation in terms of evolution difficult. To overcome these difficulties, astronomers have found alternative solutions such as classifying galaxies by spectral type~\citep{Madgwick02} or by spectro-photometric type~\citep{Zucca06}. However, direct interpretation of these results in the framework of galaxy evolution is not straightforward since galaxies move from one spectral class to another by a passive evolution of their stellar populations. A classification based on structural parameters is less sensitive to the star formation history, hence more robust to follow similar galaxies at different redshifts. 

 In the visible, progress over the last ten years have come in particular from the Hubble Deep Fields (HDF) observed with the Hubble Space Telescope. They brought observational evidence that galaxy evolution is differentiated with respect to morphological type and that a large fraction of distant galaxies have peculiar morphologies that do not fit
into the elliptical-spiral Hubble sequence \citep{Bri98, Wo03, Il06}. There is some evidence indeed that most of the stellar mass assembly occurs around $z\sim1.5-2$ (e.g. \citealp{Abraham07}). A better understanding of the physical processes that lead to the present Hubble sequence should come therefore from observations in this redshift range. In this context, near infrared observations are particularly important because the Ks-band flux at $z\sim1$ is less dependent on the recent history of star formation, hich peaks in the rest-frame UV and thus gives a galaxy type from the distribution of old stars, more closely related to the underlying total mass than optical observations.  A number of Ks-band surveys have being carried out using ground-based telescopes with different spatial coverages and limiting magnitudes (e.g. \citealp{Gard93,Mc00, Ma01}). In all cases the morphological information is poor because of the seeing-limited spatial resolution. Some morphological analyses have been performed using NICMOS on HST which is the only space instrument available in this wavelength range. All the results are on small areas involving a few tens of galaxies (e.g. \citealp{Conselice07,saracco08}) or in clusters (e.g \citealp{Zirm08}) but there are no extensive morphological classifications of field galaxies. 

In (\citealp{Huertas-Company08}, hereafter Paper I) we proposed therefore a generalization of the non-parametric morphological classification methods that uses an unlimited number of dimensions and non-linear separators, enabling us to use all the information brought by the different morphological parameters simultaneously. We showed that when applied to seeing limited data it reduces errors by more than a factor 2 compared to classical non-parametric methods, leading to a mean accuracy of $\sim$80\% of correct classifications.

 In this paper, we use this method to quantify the morphologies of $\sim$$50\,000$ galaxies based on structural parameters measured in the near infrared. Galaxies are observed in the COSMOS field with WIRCam at CFHT in Ks-band. We perform basic statistics such as redhisft and magnitude counts per morphological type and compare them to the ones obtained on the same objects using HST/ACS imaging (I-band) in order to quantify morphological k-correction effects. This classification is intended as a framework for future studies of the evolution of counts, luminosities, luminosity densities, correlation function for each morphological type over several redshift bins.
 
The paper proceeds as follows: the data set and the sample selection are presented in the next section. In Sect.~\ref{sec:morpho} we describe the technique used to quantify morphologies. In Sect.~\ref{sec:results}, we analyze the results of the classification and show the first statistics. We finally compare the results with the ones obtained with HST/ACS in Sect.~\ref{sec:hst} and discuss the effects of morphological k-correction.  

We use the following cosmological parameters throughout the paper: $H_0 = 70\,\mathrm{km\,s^{-1}\,Mpc^{-1}}$ and
$(\Omega_\mathrm{M}, \Omega_\Lambda)=(0.3,0.7)$ and the AB system for magnitudes.

\section{The data}
\label{sec:dataset}

\subsection{Description}

\begin{figure} 
 \centering 
  \resizebox{\hsize}{!}{\includegraphics{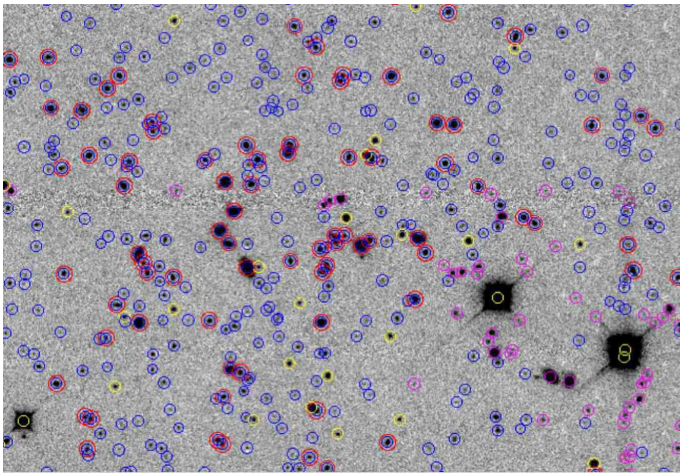}}
 \caption{$3^{'}\times2^{'}$ cutout of the observed area. Yellow circles are stars, magenta circles are masked objects, blue circles are galaxies and red circles are galaxies with computed morphology. (see text for details)} 
 \label{fig:cutout} 
 \end{figure}

The Ks-band data were taken with WIRCam \citep{thibault03} installed at CFHT in the near infrared $K_s$ band ($2.2\mu m$). The observed area is $2.65$ $deg^{2}$ and is centered in the COSMOS field \citep{Sco05}.  Images are reduced with the Terapix
pipeline\footnote{http://terapix.iap.fr} and have a pixel scale of
$0.15^{"}$ with a mean FWHM of $0.7^{"}$. 
Figure~\ref{fig:cutout} shows a cutout of the final reduced image.Data are complete up to $Ks(AB)\sim23.5$. A more detailed description of the data set can be found in McCracken et al. 2008 (in preparation). \\

The I-band data used in \S~\ref{sec:hst} are part of the COSMOS HST/ACS field \citep{Koekemoer07}. The data set consists of a contiguous 1.64 $deg^{2}$ field covering the entire COSMOS field. The Advanced Camera for Surveys (ACS) together with the F814W filter (``Broad I'') were employed. 

%More precisely, in this paper we will use the morphological catalog described in Tasca et al. 2008 (in preparation).

%These data are particularly
%interesting because Ks-band data have the advantage of probing old
%stellar populations in the rest-frame, enabling a determination of
%galaxy morphological types unaffected by recent star formation. Moreover, no space data in this wavelength range are available today.

\subsection{Building the Ks-band catalog}

\subsubsection{Detection and cleaning}

All objects having a $1.5\sigma$ signal above sky, over four contiguous pixels
are detected using \textsc{SExtractor} \citep{Ber96}.  We then performed a
cleaning task in order to separate galaxies from stars and spurious
detections. This was made using the \textsc{SExtractor} MU\_MAX and
MAG\_AUTO parameters that give the peak surface brightness above the background
and the Kron-like elliptical aperture magnitude, respectively. The distribution
of objects in this parameter space clearly defines three regions that separate extended sources from point-like or non-resolved
sources and from spurious detections. In this
separation scheme, objects with very faint magnitude and high peak
surface brightness are considered as false detections. 
The final reduced tiles have several strips where the noise is considerably higher than in the other regions of the image. Therefore, objects that fall in these regions were masked. Furthermore a mask was also applied to the objects that are too close to bright stars ($K_s<15$). The final area after having applied the mask is $2.08$ $deg^{2}$. 
We finally obtain, after cleaning and masking, $282\,122$ non-spurious sources over the whole field.

\subsubsection{Star/galaxy separation}

In order to select galaxies from the total K-selected photometric sample, we have used a number of photometric parameters to remove candidate stars, as described below. Some of the parameters that can be used are: (i) the CLASS\_STAR parameter given by \textsc{SExtractor} providing the ``stellarity-index'' for each object, (ii) the MU\_CLASS, i.e. the position in the MU\_MAX-MAG\_AUTO plane described above, reliable up to $K_s\sim20$ (iii) the $\chi^{2}$ of the SED fitting carried out during the photometric redshift estimate~\citep{Ilb06}, with templates SEDs of both stars and galaxies.

We decided to use a combination of these criteria. For objects brighter than $K_s=20$, we selected as stars objects having a $CLASS\_STAR>0.95$ and a photometric redshift lower than 0.05 or a stellar spectral class from the SED fitting in addition to the MU\_CLASS parameter. For objects with $K_S>20$ the MU\_CLASS parameter was not used and the spectral class was only used when available.  The final sample consists in $27\,343$ point-like sources and $254\,779$ galaxies.  

%Figure~\ref{fig:ncounts} shows the number counts for galaxies after removal of stars, spurious and masked objects. 

%Boundaries were drawn
%manually and a visual inspection confirms that known stars in the field
%are indeed identified as point sources.  

%(27343 point-like sources and 254779 galaxies).

%The sample is complete up to $K_{AB}\sim23$.

\begin{comment}
\begin{figure} 
 \centering 
  \resizebox{\hsize}{!}{\includegraphics{1255fig2.ps}}
 \caption{Number counts. Spurious detections and objects in masked regions have been removed. Error bars are calculated using Poisson $\sqrt{n}$ statistics. The vertical dashed line shows the magnitude limit used for morphological analysis. Results from other recent works are also plotted. }
 \label{fig:ncounts} 
 \end{figure}
\end{comment}

%\begin{itemize}

%\item sample selection for morphological analysis
%We finally cut the catalog to $K_s<21.5$ and $z_{phot}<2$. This results in a final morphological catalog of 57123 objects.

%\end{itemize}

\subsubsection{The photometric redshift catalog}

Photometric redshifts are computed in \cite{Ilbert09} using a $\chi^2$ template-fitting method. They are computed with 30 broad, intermediate, and narrow bands covering the UV (GALEX), visible-NIR (SUBARU, CFHT, UKIRT and NOAO) and mid-IR (Spitzer/IRAC). Measurement were calibrated with large spectroscopic samples from VLT-VIMOS and Keck-DEIMOS. Comparison of the derived photo-z with 4148 spectroscopic redshifts ($\Delta_z=z_s-z_p$) indicates a dispersion of $\sigma_{\Delta_z/(1+z_s)}=0.007$ at $i_{AB}<22.5$. For more details on how the redshifts are computed, please see \cite{Ilbert09}. 

Our final catalog has 198684 objects with realiable photometric redshift 
measurements.

\begin{figure} 
 \centering 
  \resizebox{\hsize}{!}{\includegraphics{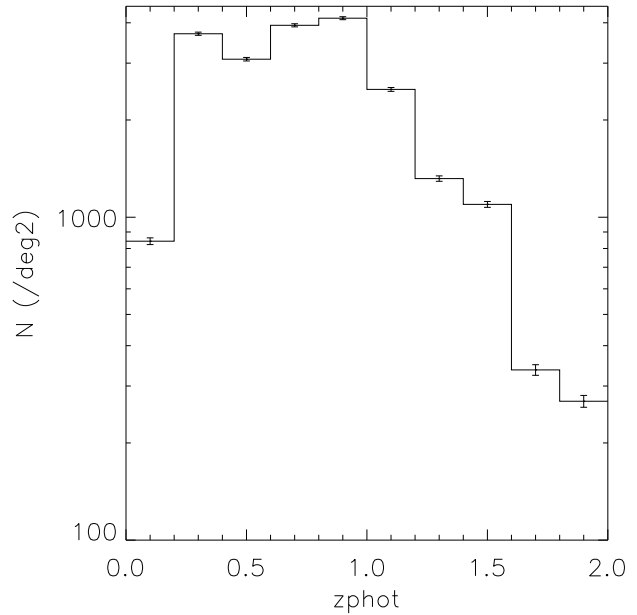}}
 \caption{Redshift distribution for the $44\,089$ analyzed galaxies. Error bars are calculated using Poisson $\sqrt{n}$ statistics.} 
 \label{fig:zphot} 
 \end{figure}

\subsubsection{The morphological catalog}
The morphological analysis is made in a subsample of the initial catalog: first we select only the galaxies which have a measured photometric redshift.  Then we cut the catalog to $K_s<21.5$ and $z_{phot}<2$. This decision is based in a visual inspection; objects fainter than 21.5 have a S/N per pixel lower than 5, so we decided not to include them in the morphological study. Simulations of those objects in fact show that the morphological classifications obtained are highly contaminated. Photometric redshift above $z\sim2$ are not reliable enough according to \cite{Ilbert09}. This selection results in a final morphological catalog of $44\,089$ galaxies. Fig~\ref{fig:zphot} shows the redshift distribution of the final catalog. 
 
\begin{figure} 
 \centering 
  \resizebox{\hsize}{!}{\includegraphics{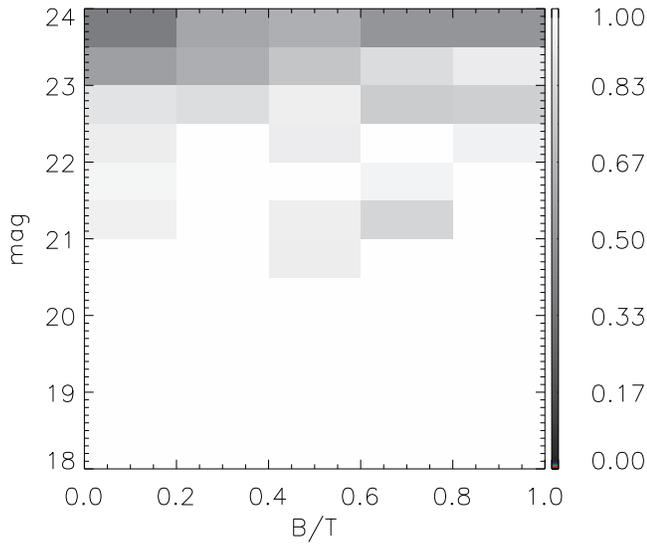}}
 \caption{Completeness for extended sources as a function of the bulge fraction ($B/T$) as assessed from a mock sample of 5000 galaxies dropped in the field images.} 
 \label{fig:comp_gal} 
 \end{figure}

\begin{figure}[h!]
\begin{center}
$\begin{array}{c@{\hspace{1in}}c}
%\multicolumn{1}{l}{\mbox{\bf (a)}} &
	%\multicolumn{1}{l}{\mbox{\bf (b)}} \\ [-0.53cm]
	\includegraphics[width=0.3\textwidth,height=0.3\textwidth]{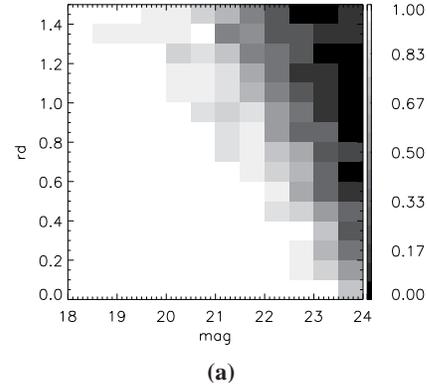} \\
	\mbox{\bf (a)} \\
	\includegraphics[width=0.3\textwidth,height=0.3\textwidth]{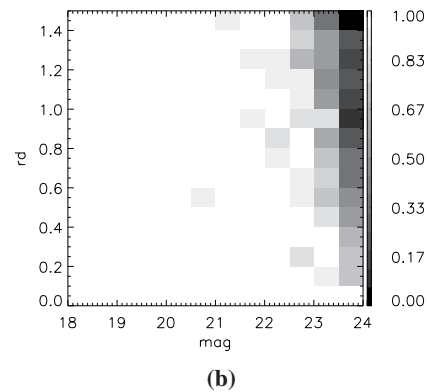} \\
       \mbox{\bf (b)}\\
%	\includegraphics[width=0.38\textwidth,height=0.3\textwidth]{huertas_fig3e.eps} &
%	\includegraphics[width=0.38\textwidth,height=0.3\textwidth]{huertas_fig3f.eps} \\ 
%	\mbox{\bf (e)} & \mbox{\bf (f)}\\
%\mbox{\bf (a)} & \mbox{\bf (b)}
	%\multicolumn{1}{l}{\mbox{\bf (b)}} \\ [-0.53cm]
	
%\mbox{\bf (c)} & \mbox{\bf (d)}

\end{array}$
\end{center}
\caption{Completeness as a function of size for simulated disks (a) and bulges (b) as assessed from a mock sample of 5000 galaxies dropped in the field images. The size is represented by the disk scale length for disks ($r_d$) and by the bulge effective radius ($r_e$) for bulges.}

\label{fig:comp_size}
\end{figure}

In order to verify precisely whether the sample is complete for all the morphological types and to quantify the selection effects, we generated $5000$ fake galaxies with exponential \citep{Fre70} and de Vaucouleurs profiles \citep{deVauc48} of different morphological types (bulge fraction uniformly distributed between 0 and 1) and with galaxy sizes uniformly distributed between $0^{"}$ and $1.5^{"}$ and dropped them in the real background images. We then tried to detect them with \textsc{SExtractor}. Results are shown in Figure~\ref{fig:comp_gal}. As we can see, the sample is complete for all the bulge fraction values up to our magnitude limit ($Ks=21.5$). We also looked at the completeness as a function of size. We generated for that purpose pure bulge and pure disk profiles with different sizes and detect them with \textsc{SExtractor} (Fig.~\ref{fig:comp_size}). As expected bulges are detected up to fainter magnitudes, however the sample is complete up to $Ks=21.5$ for both bulges and disks for sizes ranging from $0^{"}$ to $1.5^{"}$.

\section{Morphology}
\label{sec:morpho}
Galaxies in the catalog have been separated into two main morphological types (late-type and early-type) using the free available code galSVM\footnote{\url{http://www.lesia.obspm.fr/~huertas/galsvm.html}} \citep{Huertas-Company07}. By late-type we mean spiral and irregular galaxies and early-type galaxies include elliptical and lenticular types. galSVM is a non-parametric N-dimensional code based on support vector machines (SVM) that uses a training set built from a local visually classified sample. 
The employed procedure can be summarized in 4 main steps (see Paper I for more details):

\begin{enumerate}

\item Build a training set: we select a nearby visually classified sample at wavelengths corresponding to the rest-frame of the high redshift sample to be analyzed. In our case, we want to simulate Ks-band observations. We used therefore an SDSS local sample observed in the i and z bands which roughly corresponds to the rest-frame wavelengths between $z\sim1$ and $z\sim2$ for Ks-band observations. We then move the sample to the proper redshift and image quality and drop it in the real background.

\item Measure a set of morphological parameters on the sample.

\item Train a support vector based learning machine with a fraction of the simulated sample and use the other fraction to test and estimate errors.

\item Classify real data with the trained machine and correct for possible systematic errors detected in the testing step.

\end{enumerate}

\subsection{The training sample}
The most important step in obtaining the morphology with a non-parametric method is to correctly calibrate the volume filled by the data in the multi-dimensional space. This is a critical step since it will determine the decision regions that will be used to perform the classification. Indeed, galaxy morphology derivation depends on the physical properties of the galaxy (luminosity, redshift, wavelength) and on the observing conditions (background level, resolution). A suitable calibration set should consequently reproduce closely all the properties of the sample to be analyzed. One classical approach consists in visually classifying a fraction of the sample and use it as a training set to optimize boundaries \citep{Men99,Men06}. However this is not possible for seeing limited data where the resolution is too low to enable a reliable visual classification. Here, we then decide to simulate the high redshift sample from a visually classified local catalog, selected in the rest-frame color of the high redshift sample. This has three main advantages: first, it is less affected by K-correction effects, second it does not introduce any modeling effect, since the used galaxies are real and finally, the training set is built to reproduce the observing conditions and physical properties of the sample to be analyzed, but it is classified safely on well-resolved images, so it does not need to have a specially high resolution.

We use a catalog of 1319 objects from the Sloan Digital Sky Survey observed in two photometric bands (z and i) and visually classified \citep{Tasc06}. As explained in paper I, for every galaxy stamp we first generate a random pair of (magnitude, redshift) values with a probability distribution that matches the real magnitude and redshift distribution of the sample to be simulated and then we proceed in four steps: a) removal of foreground stars, b) degradation of the resolution according to a $\Lambda$CDM cosmology, c) binning to reach the desired pixel scale and d) dropping in a real background image. 

The photometric band (z or i) used to create the galaxy depends on the associated redshift. We choose the one that is closer to the rest-frame band at this given z.

We did not take into account any variation of the PSF within the field for performing the simulations. We in fact expect this variation to be small and consequently not to induce strong changes in the morphology since the analyzed galaxies are significantly larger than the PSF.

Among all the simulated objects, 450 were used as training and the remaining 869 as test.

\subsection{Classification procedure}
\label{sec:class}
We use 450 simulated galaxies as training sample. The morphological mixing is fixed to 50/50, i.e. 50\% of early-type galaxies and 50\% of late-type. Even if this is not a realistic distribution it is required to minimize the errors in the SVM classification. The classification is made in a 10-D volume with a Radial Basis Function Kernel (see PaperI for more details). The measured parameters include 6 morphological parameters: Asymmetry, Concentration (\citealp{Con00} and \citealp{Abr96} definitions), Gini \citep{Abr03}, M20 \citep{Lotz04}, Smoothness \citep{Con03}, a distance parameter (photometric redshift), a shape parameter (elongation), 2 luminosity parameters (surface brightness and magnitude). See Paper I for details on how these parameters are calculated. 

In Paper I we show, thanks to the test sample, that increasing the number of parameters of the classification never results in a degeneracy that decreases the accuracy. We realized however that when dealing with the real sample, some parameters might produce a bias if they are not realistic. This is the case of the magnitude. Indeed when doing the simulations we use the same magnitude distribution for early and late-type galaxies. This creates an artificial excess of early-type galaxies at faint magnitudes in the simulated sample that is not seen in the real world. Therefore if the magnitude is used as input parameter for the classification the number of early-type galaxies are overestimated. To avoid this bias we proceed in 2 steps: first we make a 9-D classification (without the magnitude) and then we use the measured magnitude distribution per morphological type to generate a new simulated sample with this magnitude distribution. We use this second sample to make the final 10-D classification.

\subsection{The output catalog}

Classical support vector classifiers only predict class label (i.e. early-type or late-type) but not probability information. Recently some authors \citep{Platt99,Wu04} have proposed different methods to estimate a posteriori probability, i.e. given k classes of data, for any {\bf x} the goal is to estimate $p_i=p(y=i|{\bf x}), i=1,...,k$. 
The free available package libSVM \citep{CC01a} implements the method described in \cite{Wu04}. Therefore we added this new feature to our classification output:
{\bf we} associate to every galaxy in the morphological catalog, a class label and a probability of belonging to the given class. Since we are dealing with a 2-class problem, the probability p(galaxy=early-type)=1-p(galaxy=late-type). In the following, we use this parameter to assess the accuracy of our classification.

\subsection{Accuracy}
\label{sec:accuracy}
As shown in paper I, one of the main advantages of the employed method is that the reliability of the classifications can be quantified using a test sample simulated in the same way as the training one.

Here we use the probability as the main estimator. We first looked at the evolution of the correct classifications as a function of different probability thresholds. Results are shown in table~\ref{tbl:accu_comp}. As we can see, there is a clear correlation between the probability threshold and the number of correct identifications: the accuracy clearly increases when the considered probability is higher. If we select only objects with a probability between $0.5$ and $0.6$ the mean accuracy is only around $58\%$. However, objects with probabilities greater than 0.8 are classified with nearly $90\%$ accuracy. The contamination is around $\sim$20\% for the whole sample ($p>0.5$).

\begin{comment}
\begin{figure}[h!] 
 \centering 
  \resizebox{\hsize}{!}{\includegraphics{correct_vs_proba.eps}}
 \caption{Correlation between } 
 \label{fig:correct_vs_proba} 
 \end{figure}
 \end{comment}

\begin{figure*}[h!t!]
\begin{center}
$\begin{array}{c@{\hspace{1in}}c}
%\multicolumn{1}{l}{\mbox{\bf (a)}} &
	%\multicolumn{1}{l}{\mbox{\bf (b)}} \\ [-0.53cm]
	\includegraphics[width=0.38\textwidth,height=0.3\textwidth]{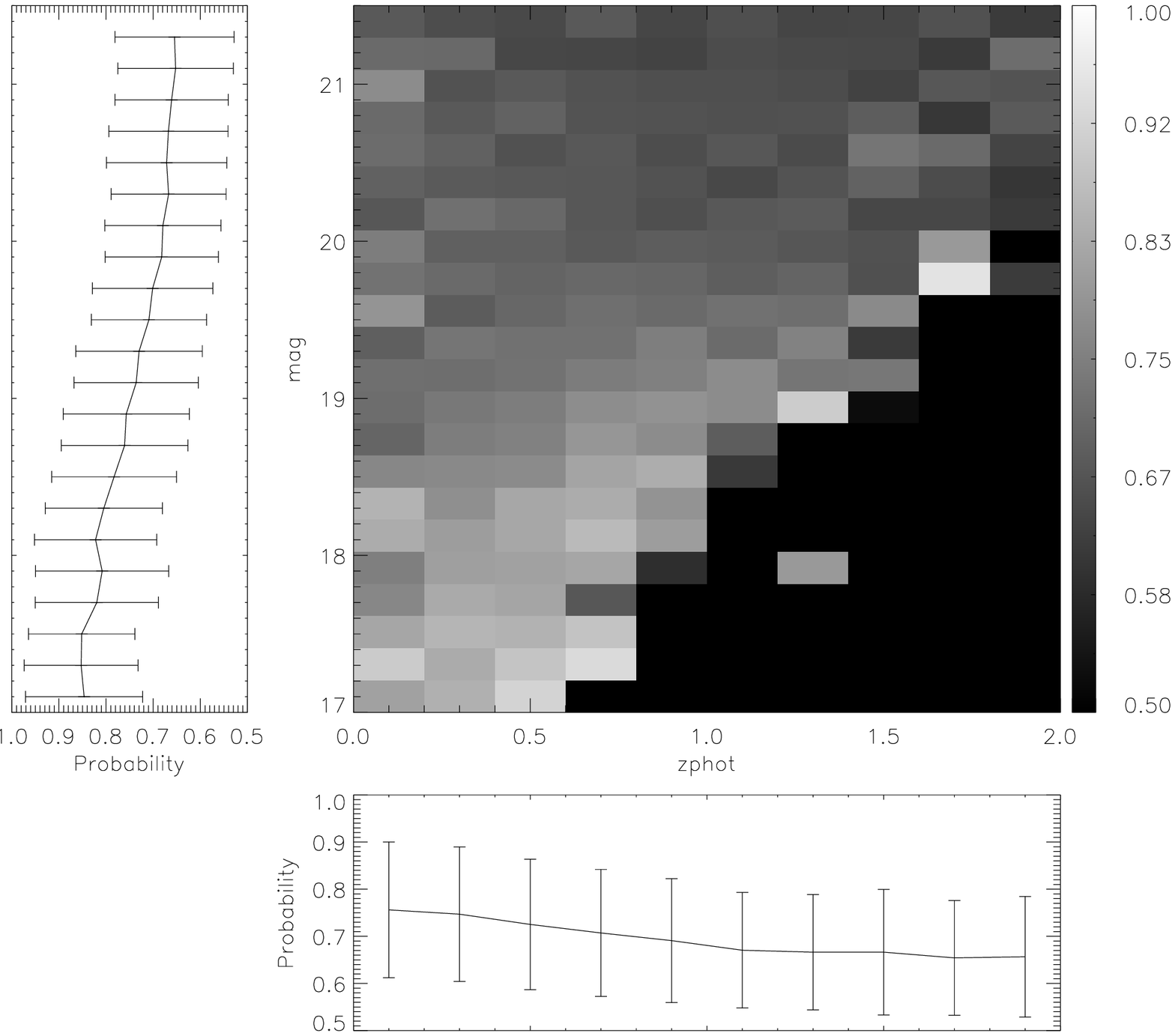} &
	\includegraphics[width=0.38\textwidth,height=0.3\textwidth]{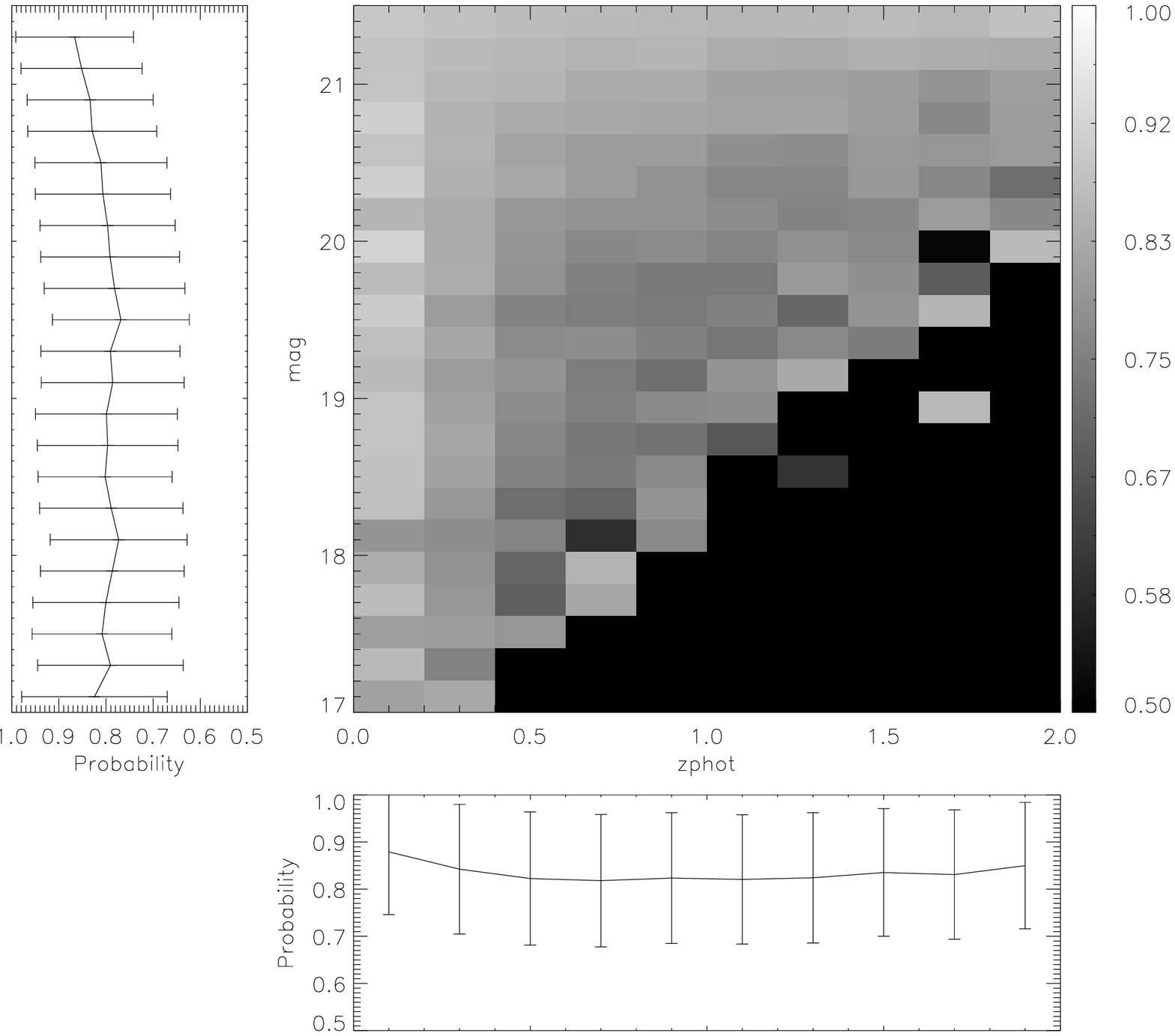} \\
\mbox{\bf (a)} & \mbox{\bf (b)}\\
	\includegraphics[width=0.38\textwidth,height=0.3\textwidth]{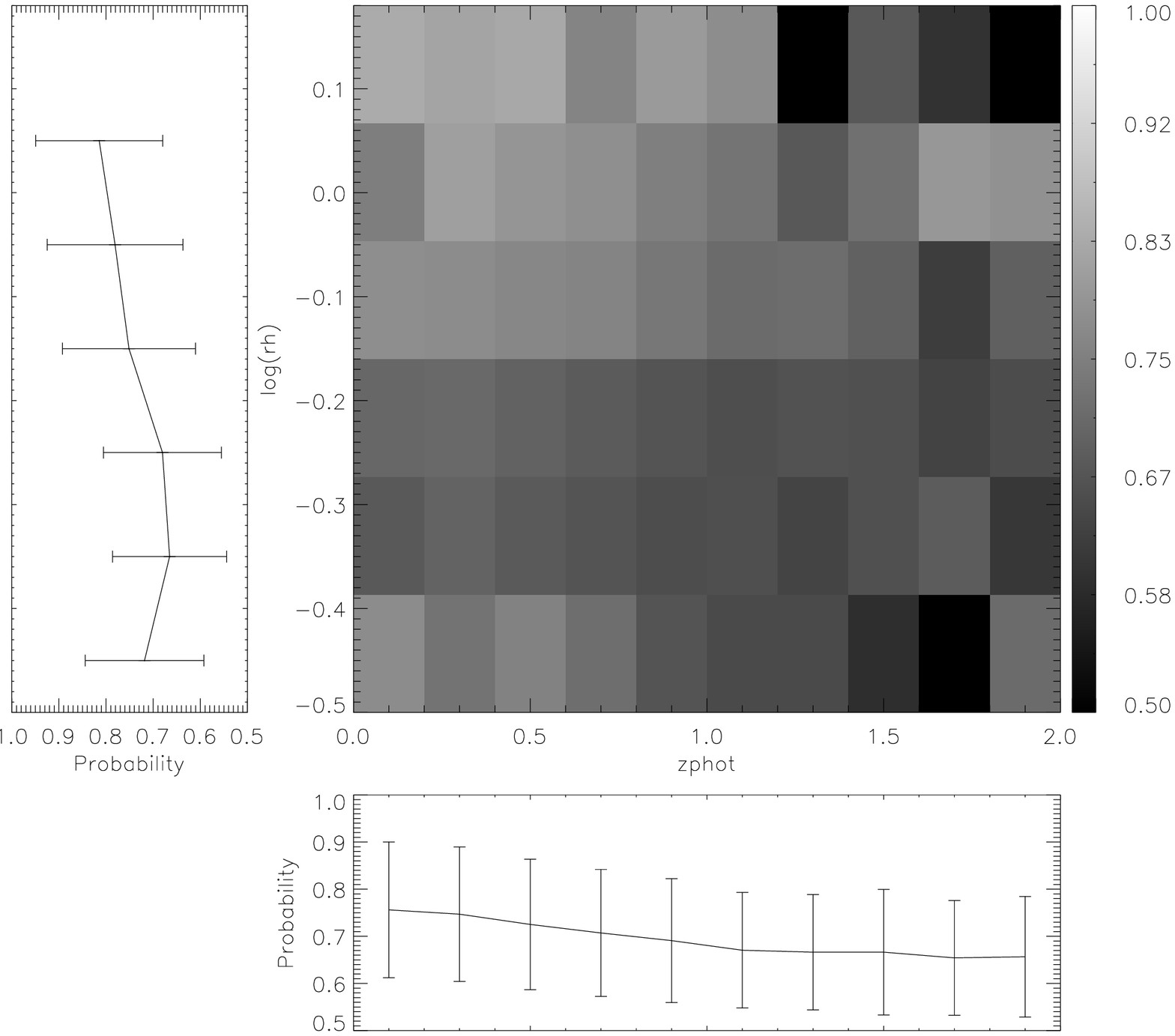} &
	\includegraphics[width=0.38\textwidth,height=0.3\textwidth]{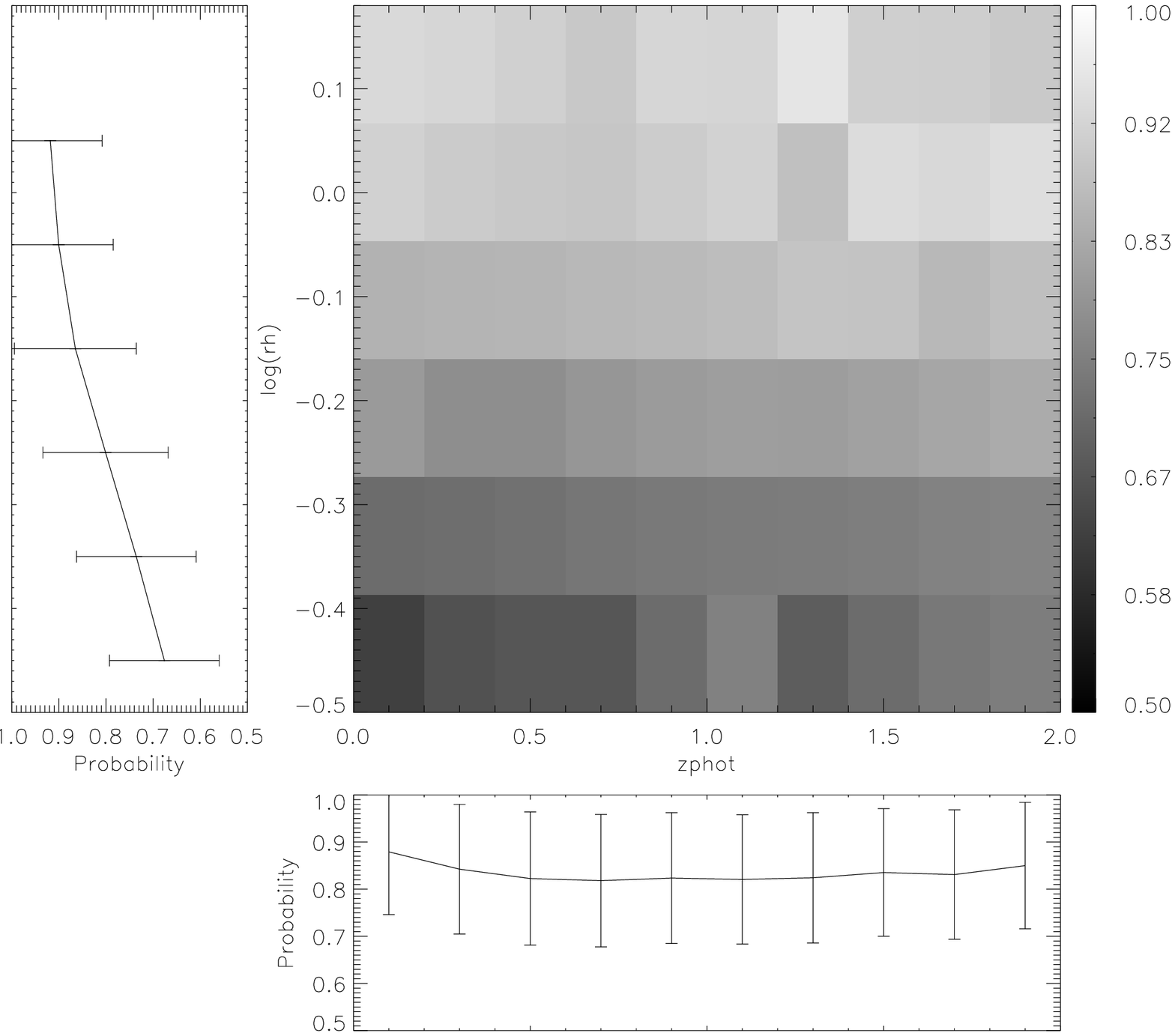} \\ 
	\mbox{\bf (c)} & \mbox{\bf (d)}\\
	\includegraphics[width=0.38\textwidth,height=0.3\textwidth]{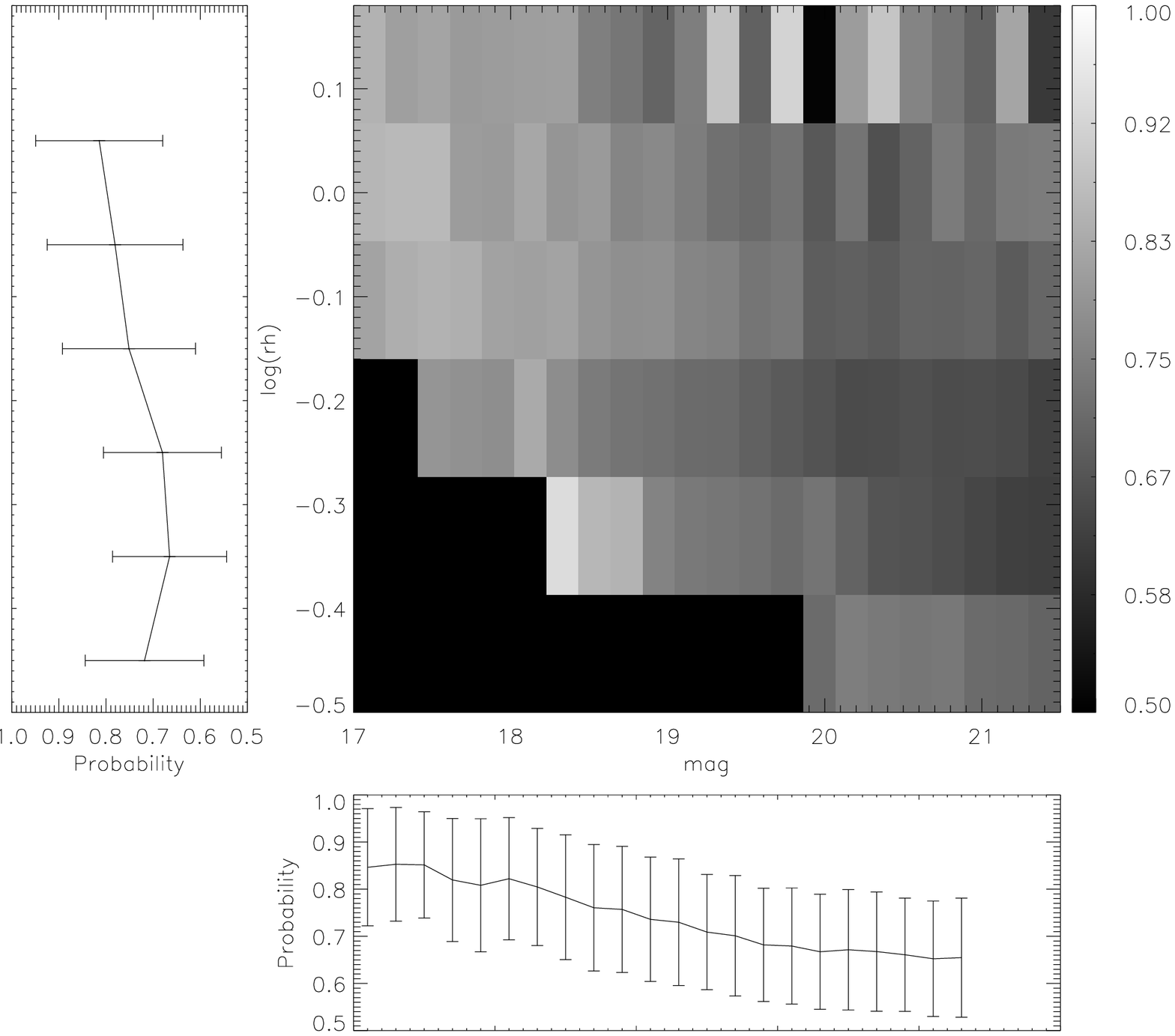} &
	\includegraphics[width=0.38\textwidth,height=0.3\textwidth]{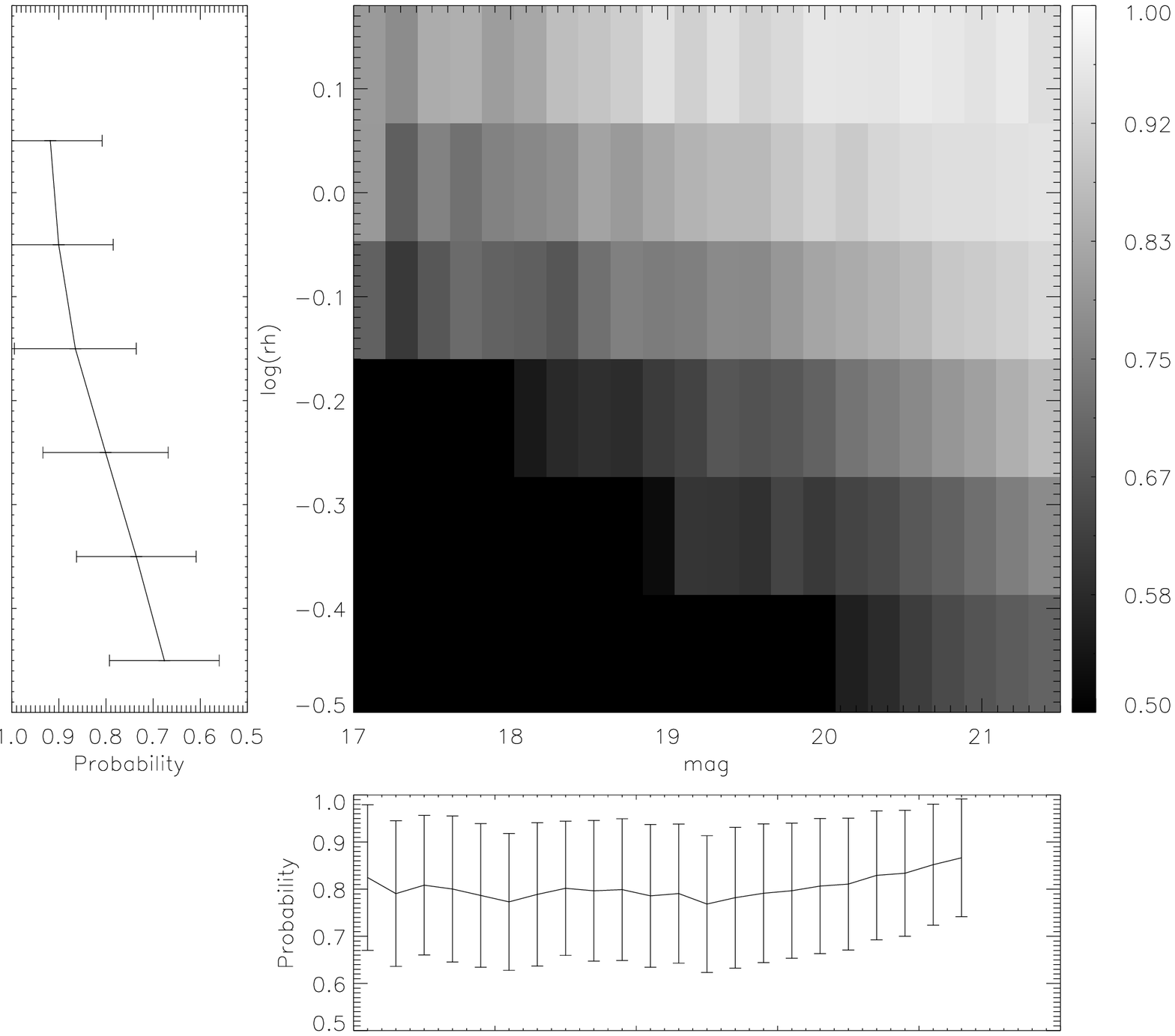} \\ 
	\mbox{\bf (e)} & \mbox{\bf (f)}\\
%\mbox{\bf (a)} & \mbox{\bf (b)}
	%\multicolumn{1}{l}{\mbox{\bf (b)}} \\ [-0.53cm]
	
%\mbox{\bf (c)} & \mbox{\bf (d)}

\end{array}$
\end{center}
\caption{Bi-dimensional maps of the mean probabilities for different redshift, magnitude and size bins of the real sample. The size is represented by the half light radius as measured by SExtractor. The plots show the mean probability distribution as a function of a single parameter. Error bars show the dispersion of the probability distribution. On the left column we show the probability maps for early-type objects and on the right the maps for late-type objects. The black cells indicate that there are no objects in the considered bin. }
\label{fig:proba_maps}
\end{figure*}

 This relations between the probability parameter and the success rate enable the use of the probability  as an estimate of the classification accuracy as a function of physical parameters of the galaxies directly on the real sample. We decided to represent 2D maps of the mean probability values (Fig.~\ref{fig:proba_maps}) for different magnitude, redshift and size bins. We observe several interesting trends:
 \begin{itemize}
 \item Globally, late-type galaxies have higher probabilities for all redshift, magnitude and size values. This indicates that late-type objects are easily isolated. It is probably a consequence of the ellipticity parameter, used in the classification procedure. Indeed, objects with high ellipticity are identified as late-type galaxies with high probability so that the mean probability increases.
 \item  As expected there is a clear trend with size for both morphological types, small objects ($r_{half}<0.6^{"}$) have lower probabilities ($p\sim0.7$) while large ones ($r_{half}>0.6^{"}$) have higher values ($p>0.8$) (Fig.~\ref{fig:proba_maps}~c,d).
 \item There is also a trend with redshift, especially for early-type objects: below $z\sim1$ the mean probability is around $p\sim0.75$ and it decreases to $p\sim0.7$ for $z>1$. The number of galaxies with low probability ($p\sim0.6$) also increases (Fig.~\ref{fig:proba_maps}~a,b).
 \item Finally, the magnitude is also important for determining the quality of the classification. Above $K_s\sim20$ the mean probability for early-type objects is around $p\sim0.7$ (Fig.~\ref{fig:proba_maps}~a,e). Interestingly, this trend does not appear for the late-type population (Fig.~\ref{fig:proba_maps}~b,f). This is probably a consequence of the way the training sample is built (\S~\ref{sec:class}): at faint magnitudes, the number of early-type galaxies is low so the machine tends to find late-type galaxies with higher probability. 
  \end{itemize} 

\begin{table*}
\begin{center}

\caption{Accuracy of the classifications based on the test sample for a 10 parameters SVM training (C, A, G, Ell, S, M20, C2, SB, z, mag) as a function of probability. The table shows the fraction of correct classifications when a probability bin $[P_{min},P_{max}]$ is considered. Numbers between brackets give the number of objects per bin. Top: early-type galaxies, bottom: late-type galaxies.}
\begin{tabular}{c|c|c|c|c|c|c|c|c|c|c|c|}
\hline\hline\noalign{\smallskip}
$P_{max}\rightarrow$ & 0.5 & 0.55 & 0.6 & 0.65 & 0.7 & 0.75 & 0.8 & 0.85 & 0.9 & 0.95 & 1.0 \\
 $P_{min}\downarrow$   & & & & & & & & & & &\\ 
\noalign{\smallskip}\hline\noalign{\smallskip}
0.5 & --&--& 0.56 (45) &0.55 (63) & 0.53 (82) &  0.6 (111) &  0.66 (157) & 0.7 (195) & 0.73 (236) & 0.76 (315) & 0.78 (352) \\
0.55 & --&--&--& 0.58 (46) &0.56 (65) & 0.62 (94) &  0.68 (140) & 0.72 (178) & 0.76 (219) & 0.80(298) & 0.82 (335) \\
0.6 & --&--&--& -- &0.52 (37) & 0.62 (66) &  0.70 (112) & 0.74 (150) & 0.78 (191) & 0.83(270) & 0.84 (307) \\
0.65 & --&--&--& -- & -- & 0.63 (48) &  0.75 (94) & 0.78 (132) & 0.82 (173) & 0.87(252) & 0.87 (289) \\
0.7 & --&--&--& -- & -- & -- &  0.85 (75) & 0.86 (113) & 0.9 (154) & 0.92(233) & 0.92 (270) \\
0.75 & --&--&--& -- & -- & -- &  -- & 0.89 (84) & 0.92 (125) & 0.94(204) & 0.94 (241) \\
0.8 & --&--&--& -- & -- & -- &  -- & -- & 0.93 (79) & 0.96(158) & 0.94 (195) \\
0.85 & --&--&--& -- & -- & -- &  -- & -- & -- & 0.98(120) & 0.96 (157) \\
0.9 & --&--&--& -- & -- & -- &  -- & -- & -- & -- & 0.95 (116) \\

\noalign{\smallskip}\hline
\end{tabular}

\begin{tabular}{c|c|c|c|c|c|c|c|c|c|c|c|}
\hline\hline\noalign{\smallskip}
$P_{max}\rightarrow$& 0.5 & 0.55 & 0.6 & 0.65 & 0.7 & 0.75 & 0.8 & 0.85 & 0.9 & 0.95 & 1.0 \\
 $P_{min}\downarrow$   & & & & & & & & & & &\\     
\noalign{\smallskip}\hline\noalign{\smallskip}
0.5 & --&--& 0.58 (39) &0.62 (68) & 0.62 (82) &  0.64 (114) &  0.67 (155) & 0.7 (189) & 0.72 (224) & 0.74 (271) & 0.75 (309) \\
0.55 & --&--&--& 0.68 (53) &0.66 (67) & 0.68 (99) &  0.70 (140) & 0.73 (174) & 0.74 (209) & 0.76(256) & 0.77 (294) \\
0.6 & --&--&--& -- &0.68 (46) & 0.70 (87) &  0.70 (116) & 0.74 (150) & 0.75 (185) & 0.78 (232) & 0.78 (270) \\
0.65 & --&--&--& -- & -- & 0.74 (73) &  0.75 (94) & 0.75 (121) & 0.76 (156) & 0.78 (203) & 0.79 (241) \\
0.7 & --&--&--& -- & -- & -- &  0.85 (75) & 0.78 (107) & 0.78 (142) & 0.81(189) & 0.82 (227) \\
0.75 & --&--&--& -- & -- & -- &  -- & 0.80 (75) & 0.80 (110) & 0.83 (157) & 0.83 (195) \\
0.8 & --&--&--& -- & -- & -- &  -- & -- & 0.84 (69) & 0.86(116) & 0.86 (154) \\

0.85 & --&--&--& -- & -- & -- &  -- & -- & -- & 0.85 (82) & 0.85 (120) \\
0.9 & --&--&--& -- & -- & -- &  -- & -- & -- & -- & 0.87 (85) \\

\noalign{\smallskip}\hline
\end{tabular}
\label{tbl:accu_comp}
\end{center}
\end{table*}

How can we take into account these trends for correcting our classification and performing statistical analysis? The simplest way seems to establish \emph{probability thresholds} but, what galaxies do we keep when selecting objects with a given probability or, in other words,  do we introduce any bias when selecting a sample with a given probability cut? 

\begin{figure*}[h!t!]
\begin{center}
$\begin{array}{c@{\hspace{1in}}c}
%\multicolumn{1}{l}{\mbox{\bf (a)}} &
	%\multicolumn{1}{l}{\mbox{\bf (b)}} \\ [-0.53cm]
	\includegraphics[width=0.3\textwidth,height=0.3\textwidth]{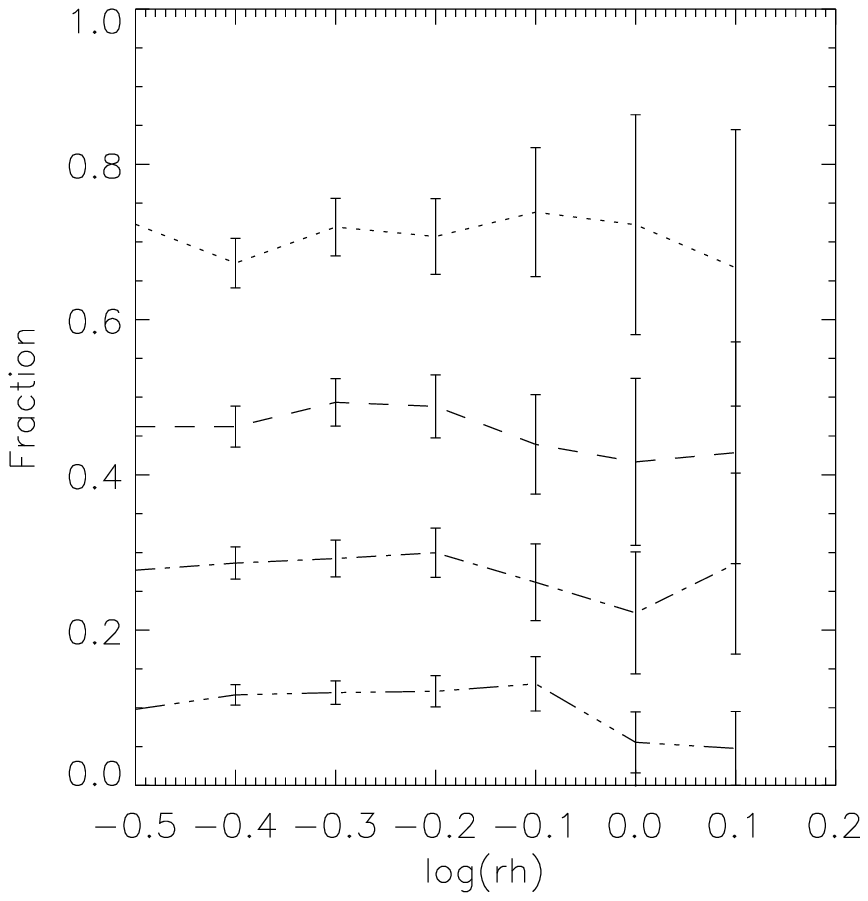} &
	\includegraphics[width=0.3\textwidth,height=0.3\textwidth]{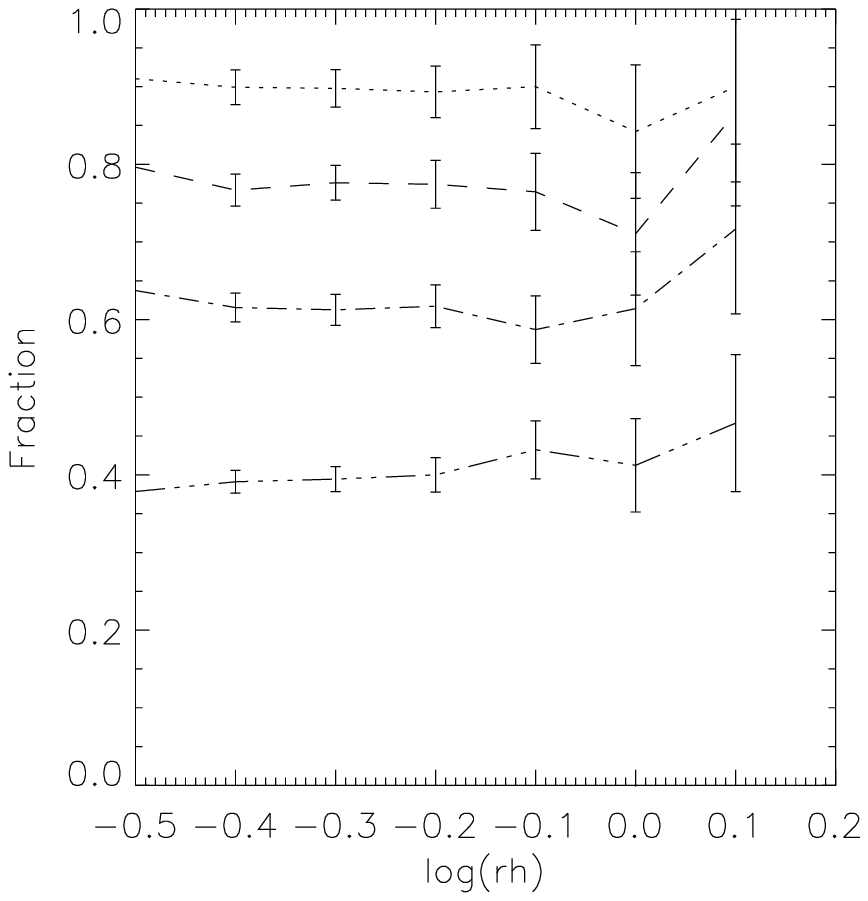} \\
	
\mbox{\bf (a)} & \mbox{\bf (d)} \\
	\includegraphics[width=0.3\textwidth,height=0.3\textwidth]{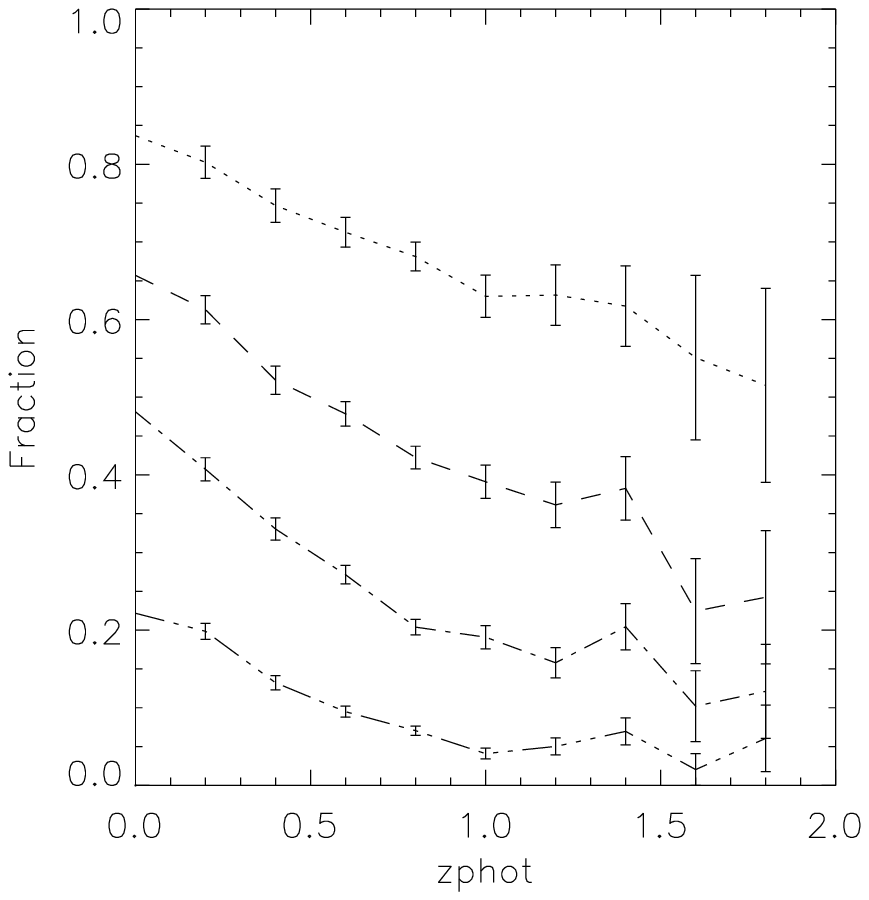} &
	\includegraphics[width=0.3\textwidth,height=0.3\textwidth]{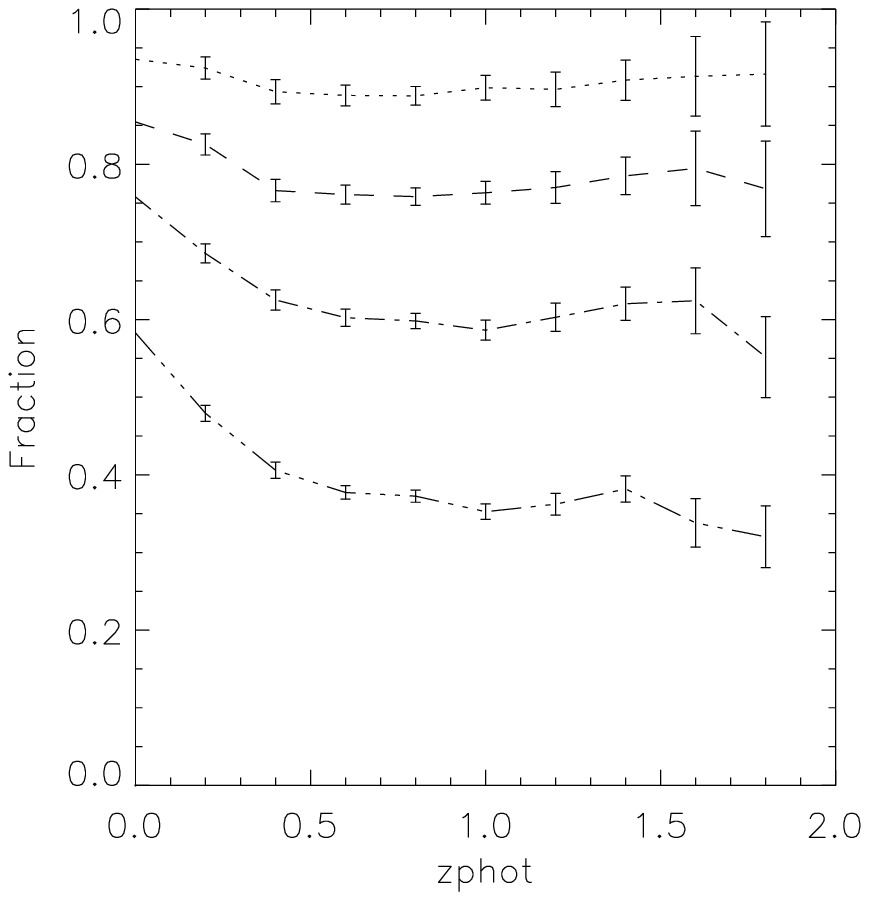} \\		
	\mbox{\bf (b)} & \mbox{\bf (e)} \\
	\includegraphics[width=0.3\textwidth,height=0.3\textwidth]{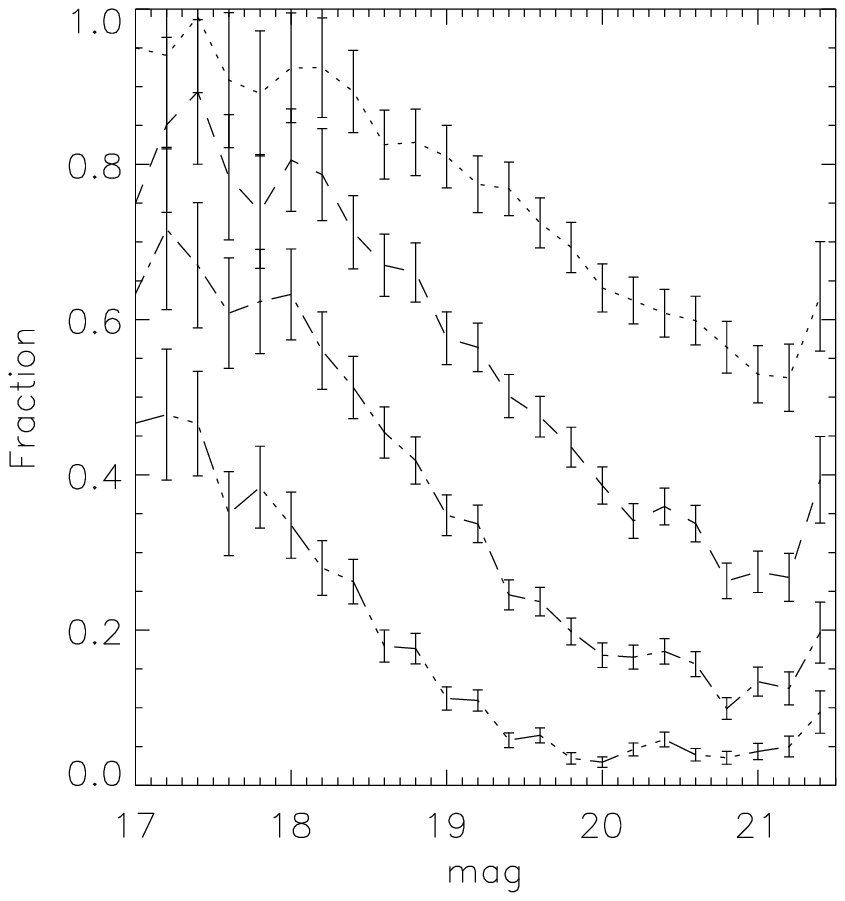} &
	\includegraphics[width=0.3\textwidth,height=0.3\textwidth]{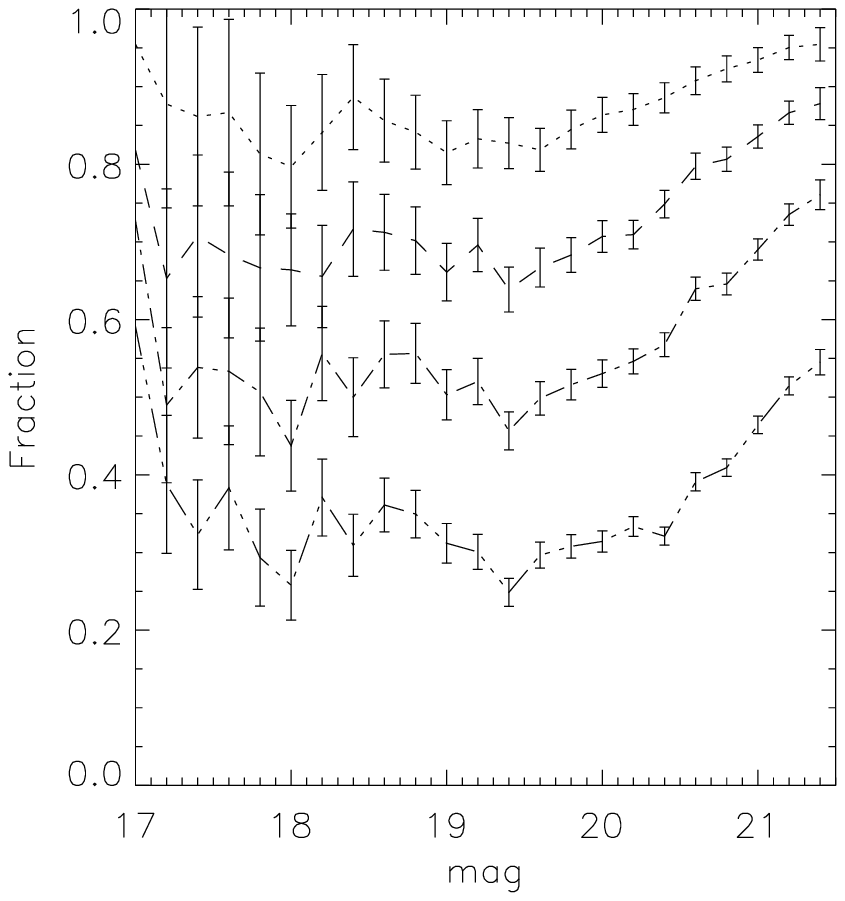} \\		
	\mbox{\bf (c)} & \mbox{\bf (f)} \\

%	\includegraphics[width=0.38\textwidth,height=0.3\textwidth]{huertas_fig3e.eps} &
%	\includegraphics[width=0.38\textwidth,height=0.3\textwidth]{huertas_fig3f.eps} \\ 
%	\mbox{\bf (e)} & \mbox{\bf (f)}\\
%\mbox{\bf (a)} & \mbox{\bf (b)}
	%\multicolumn{1}{l}{\mbox{\bf (b)}} \\ [-0.53cm]
	
%\mbox{\bf (c)} & \mbox{\bf (d)}

\end{array}$
\end{center}
\caption{Sample completeness as a function of redshift . Left column are early-type galaxies and right column late-type. (a)-(d) size (b)-(e) redshift and (c)-(f) magnitude for different probability cuts.} Dotted line: galaxies with $p>0.6$, dashed line: galaxies with $p>0.7$, dashed-dotted line: galaxies with $p>0.8$ and dashed line with three dots: galaxies with $p>0.9$. Error bars are calculated using Poisson $\sqrt{n}$ statistics.
\label{fig:complete}
\end{figure*}

To answer this question we selected objects with 4 probability thresholds ($p>0.6$, $p>0.7$, $p>0.8$ and $p>0.9$) and examined the completeness of the selected sample as a function of the magnitude, the redshift, the size and the morphological type. Figure ~\ref{fig:complete} shows the results. If no biases were introduced all the lines should be \emph{flat} (i.e ``we keep all the galaxies in the same way''). We observe in Fig.~\ref{fig:complete} that this is not the case. This is seen in particular when looking at the completeness as a function of size and morphological type: large objects are preferred as expected and early-type objects are more penalized. 

Therefore a sample selected by performing a probablity cut should not be used for global statistical analysis. It can be used however to select particular class of objects for which a very good accuracy is needed. In the next section we show however a way of using the information brought by probabilities for statistical purposes.

\begin{comment}
\begin{tabular}{c|cc|cc|}
\hline\hline\noalign{\smallskip}

 & \multicolumn{2}{c|}{$p>0.5$} & 
            \multicolumn{2}{c|}{$p>0.6$} \\           
           
 & E-T & L-T & E-T & L-T  \\
\noalign{\smallskip}\hline\noalign{\smallskip}
Vis. E-T &  0.78 (352) & 0.25 (104) &0.83 (307) & 0.20 (75)\\
Vis. L-T &   0.22 (96) & 0.75 (309) & 0.17 (62) & 0.80 (270) \\
\noalign{\smallskip}\hline
\end{tabular}

\begin{tabular}{c|cc|cc|}
\hline\hline\noalign{\smallskip}

 & \multicolumn{2}{c|}{$p>0.7$} & 
            \multicolumn{2}{c|}{$p>0.8$} \\

 & E-T & L-T & E-T & L-T  \\
\noalign{\smallskip}\hline\noalign{\smallskip}
Vis. E-T &   0.91 (270) &0.18 (53) & 0.94 (195) & 0.14 (27) \\
Vis. L-T &   0.09 (26) &0.82 (227)& 0.06 (13) & 0.86 (154) \\
\noalign{\smallskip}\hline
\end{tabular}
\end{center}

\end{table}  
\end{comment}

\subsubsection{Best estimator}

Considering that we have N galaxies with a probability $p_k$ of being in class A (e.g. early-type or late-type), we can define two functions to estimate the number of objects in a given class.

The first one (hereafter \emph{counts estimator}) uses a function
\[ Z_k = \left\{ \begin{array}{ll}
0 & \mbox{if $p_k< 0.5$}\\
1 & \mbox{if $p_k>0.5$}. \end{array} \right. \]
Then the number of objects in class A is simply: $N_A=\sum{Z_k}$. Objects with a probability greater than 0.5 of belonging to a given class are simply added. With this estimator we consider therefore in the same way a galaxy with $p=0.51$ and one with $p=0.99$; it is just added to the corresponding class and the probability is ignored. 

The second one (hereafter \emph{probability estimator}) tries to make use of the information contained in the probability parameter. It has been shown in the previous section that galaxies with higher probabilities have lower classification errors. However making arbitrary probability cuts introduces biases in the global statistics in the sense that not all the objects are removed uniformly. It would be interesting therefore to find a way of using the information contained in this parameter without introducing biases.  For that purpose, we define a random variable 
\[ Y_k = \left\{ \begin{array}{ll}
0 & \mbox{with a probability $1-p_k$}\\
1 & \mbox{with a probability $p_k$.} \end{array} \right. \]
Then we can estimate the number of objects in class A as the mathematical expectation of this variable:
$E(Y)=\sum{E(Y_k)}=\sum{p_k}$
and the 1-$\sigma$ error on the number as the square root of the variance of the variable:
$Var(Y)=E(X^2)-E(X)^2=\sum{p_k(1-p_k)}$. This way, a galaxy with a probability $p=0.51$ of being in class A counts as 0.51 in this class but also as 0.49 in class B.

Therefore, the larger the number of galaxies with probability values close to 0.5, the more the differences between the two estimators will be important. As a matter of fact, if all the probabilities are equal to one, both estimators give the same results whereas if all the probabilities are equal to 0.5, the probability estimator gives half the value of the count estimator. In this sense the comparison of the results furnished by the two estimators is a kind of measure of the classification accuracy.

\subsubsection{Errors due to the training set}
The estimator presented above takes into account the information brought by the probability parameter to estimate at best the number of galaxies of each morphological type and correct from misclassifications. However, there is another source of error which has to be considered. It is related to the training set itself. Indeed, even if the training sample is built to reproduce at best the parameters of the real one it can contain errors or there can exist galaxies which are not well represented in the parameter space. In order to estimate these errors, we performed Monte Carlo simulations: we randomly removed elements from the training sample and generated therefore multiple training samples with fewer galaxies. We then used these samples to train different machines and classify the real sample. The generated samples have similar properties than the original one but do not fill the parameter space in the same way. They can be used consequently to estimate the effect in the final classification of \emph{missing objects} in the parameter space. The differences found in the classifications are then employed to estimate a kind of \emph{confidence region} of our classification scheme in the following sections. 

\section{Results and discussion}
\label{sec:results}
\subsection{Global statistics}

We find 33291($\sim$75\%) late-type galaxies and 10798 ($\sim$25\%) early-type galaxies with the counts estimator and 30711.7$\pm$1952 (70\%$\pm5$\%) and 13376.3$\pm$2014 (30\%$\pm5$\%) early-type galaxies with the probability estimator. In the next sections we try to locate more precisely the differences.

\subsection{Number counts}
\label{sec:distributions}
\begin{figure}[h!] 
 \centering 
  \resizebox{\hsize}{!}{\includegraphics{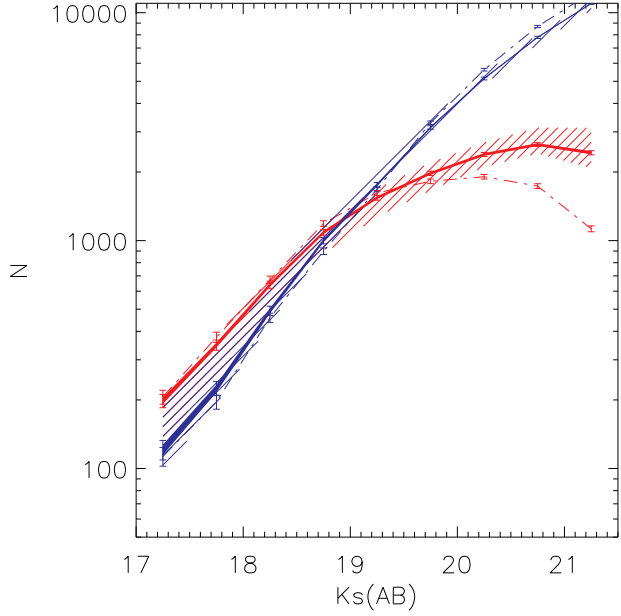}}
 \caption{Number counts per morphological type for the $44\,089$ analyzed galaxies. The solid line shows the results obtained with the probability estimator; the line width indicates the $1-\sigma$ confidence band estimated from the probability distribution and the dashed region is the confidence region deduced from Monte Carlo simulations of the training sample (see text for details). The dashed line shows the results obtained with the counts estimator. Red lines stand for early-type galaxies and blue lines for late-type. Error bars are calculated using Poisson $\sqrt{n}$ statistics.} 
 \label{fig:mag_dist} 
 \end{figure}

Figure~\ref{fig:mag_dist} shows the number counts per morphological type using the two estimators described above. As expected, early-type objects dominate the bright end of the magnitude distribution  while late-type objects are more frequent at the faint-end. The effect of using the probability estimator is seen clearly in the faint elliptical population ($K_s>20$). In particular the number of faint early-type galaxies increases when using the probability estimator. This reflects that at faint magnitudes early-type galaxies are classified with lower probabilities as seen from the 2D maps (Fig:~\ref{fig:proba_maps}). The use of the probability parameter in performing the counts enables to correct for the classification errors by adding galaxies that fell in the late-type class but had a significant probability of being elliptical.  

\subsection{Morphological evolution}
\begin{figure}[h!] 
 \centering 
  \resizebox{\hsize}{!}{\includegraphics{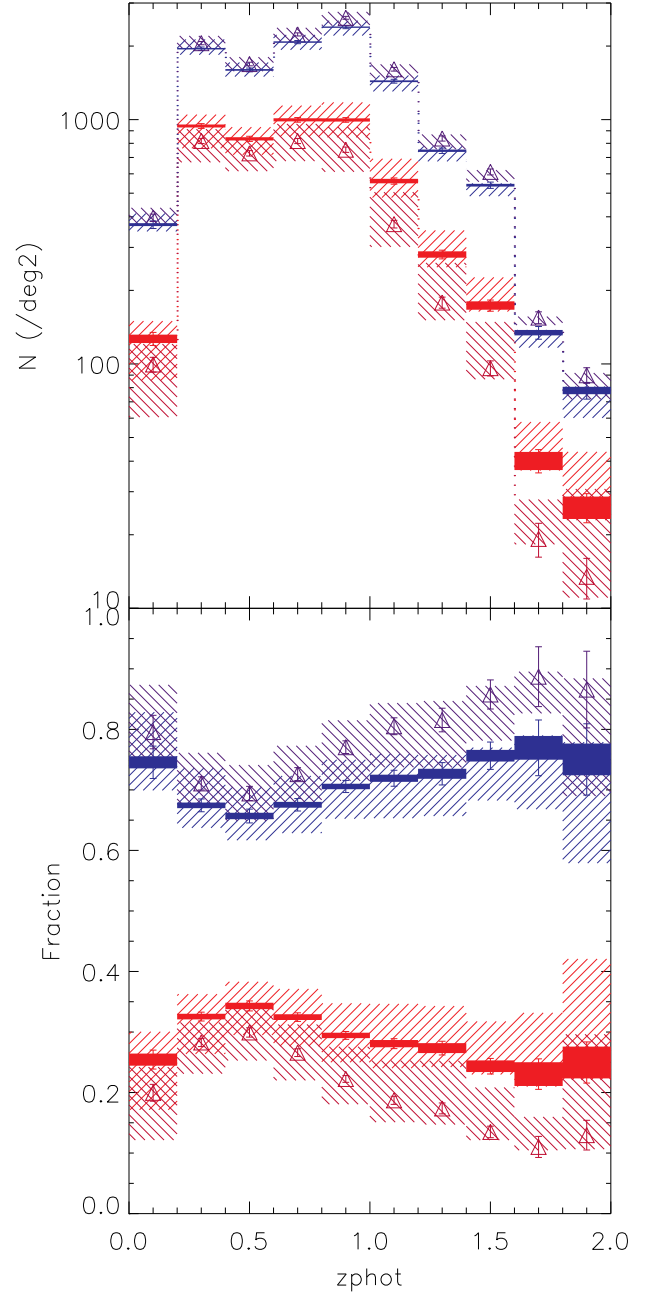}}
 \caption{Redshift distribution per morphological type for the $44\,089$ analyzed galaxies. The solid line shows the results obtained with the probability estimator; the line width indicates $1-\sigma$ confidence band estimated from the probability distribution. Empty triangles show the results obtained with the counts estimator. Dashed regions for both estimators are the confidence regions deduced from Monte Carlo simulations of the training sample (see text for details).  Red lines stand for early-type galaxies and blue lines for late-type. Error bars are calculated using Poisson $\sqrt{n}$ statistics.} 
 \label{fig:z_dist} 
 \end{figure}

Figure~\ref{fig:z_dist} shows the morphological mixing evolution up to $z\sim2$. Both estimators reveal an increase of the early type fraction from $z\sim2$, however the effect is less important when taking into account the probability. In fact, with the probability estimator we find 21.9\%$\pm$8\% early-type objects at $z\sim2$ while the local fraction is 32.0\%$\pm$5\%. The \emph{counts estimator} predicts a fraction of $\sim$15\% at $z\sim1.5$. Considering the probability helps therefore to correct from the incompleteness of the early-type population at high z due to the lower probability values. 

This variation in the early-type population is a well-known effect which has been detected using rest-frame morphologies from HST (e.g. \citealp{Bri98,Cassata05}) and probably reflects the building-up of the red sequence from late-type systems. It could be argued however that this effect was a consequence of morphological k-correction since at higher redshifts we probe the UV galaxy emission, i.e. young stellar populations. The fact that we still observe this trend with NIR data (which probe older stellar populations) seems to point that this is a real effect and not a morphological k-correction effect. It is important to point out that the fraction found here at $z\sim1.5$ is significantly higher than the one obtained in \cite{Cassata05, arn07} or \cite{Abraham07}. This short evolution could be explained by a selection effect. The $K<21.5$ selected sample does not probe the same galaxy populations at $z\sim1.5$ and $z\sim0$: at $z\sim1.5$, only the high mass end of the K-band luminosity function is sampled whereas at $z\sim0$, the LF is sampled over several magnitudes. This will be investigated in a forthcoming paper in which we will use stellar mass estimates to build a volume limited sample. In this paper we want to focus on the morphological k-correction effects and for that purpose a magnitude-limited sample is enough.  The following sections are therefore focused on this important point.

%Can these differences be a consequence of band shifting? This is a crucial point of this study and the main reason for performing morphological analysis in the NIR. 

\begin{comment}
\begin{figure*}[h!]
\begin{center}
$\begin{array}{c@{\hspace{1in}}c}
%\multicolumn{1}{l}{\mbox{\bf (a)}} &
	%\multicolumn{1}{l}{\mbox{\bf (b)}} \\ [-0.53cm]
	\includegraphics[width=0.38\textwidth,height=0.3\textwidth]{z_dist_vs_acs_44089.ps} &
	\includegraphics[width=0.38\textwidth,height=0.3\textwidth]{mag_dist_vs_acs_44089.ps} \\
\mbox{\bf (a)} & \mbox{\bf (b)}\\
%	\includegraphics[width=0.38\textwidth,height=0.3\textwidth]{huertas_fig3e.eps} &
%	\includegraphics[width=0.38\textwidth,height=0.3\textwidth]{huertas_fig3f.eps} \\ 
%	\mbox{\bf (e)} & \mbox{\bf (f)}\\
%\mbox{\bf (a)} & \mbox{\bf (b)}
	%\multicolumn{1}{l}{\mbox{\bf (b)}} \\ [-0.53cm]
	
%\mbox{\bf (c)} & \mbox{\bf (d)}
\end{array}$
\end{center}
\caption{Redshift and magnitude distributions per morphological type for 44089 galaxies. The distributions are estimated with two different estimators (see text for details)}
\label{fig:svm_dist}
\end{figure*}
\end{comment}

%\subsection{Early-type objects at $z\sim1.5$}
%\label{sec:early}

\section{Investigating the morphological k-correction effect}
\label{sec:hst}

In the previous section, we have obtained a morphological classification from NIR imaging. A crucial point is to understand the differences (if there are some) between morphologies quantified in the visible and the ones computed here. 
We perform for that purpose a match between the Ks selected morphological catalog and the morphologies measured from HST/ACS data (I-band) in an independent way. 

For the classification of the I-band sample we use a similar method as for the K-band sample: we train a 5-D support vector machine (gini, concentration, asymmetry and ellipticity). As shown in Paper I, this is largely enough for a well-resolved sample like the ACS one. Since ACS galaxies are better resolved than WIRCam ones, the training sample is not built from an SDSS sample but by visually classifying $500$ real ACS galaxies. The way the visual classification is performed is fully described in Tasca et al. 2008 (in preparation). 
%The algorithm used to morphologically classify galaxies in the I-band is described in Tasca et al. 2008 (in preparation) and we refer the reader to that paper for details on the method.

%Here it is enough to stress that the high resolution of ACS images combined 
%with a depth of I$\sim 27$  allow for accurate morphological measurements well further
%the limits imposed by this study.

 We match $32\,171$ objects of our K-selected sample. The remaining objects were too faint in the ACS data ({\bf $I>24$}) to perform reliable morphological classifications.
The global morphological mixing from HST/ACS data is in good agreement with the one obtained with WIRCam for the same objects: 9163 (I-band), 10458.8 (Ks-band) early-type galaxies and 25002 (I-band), 23706.2 (Ks-band) late-type galaxies respectively. However, where are the differences localized precisely?

\subsection{One-to-one comparison}

\begin{figure}[h!]
\begin{center}
%$\begin{array}{c}
%\multicolumn{1}{l}{\mbox{\bf (a)}} &
	%\multicolumn{1}{l}{\mbox{\bf (b)}} \\ [-0.53cm]
	 \resizebox{\hsize}{!}{\includegraphics{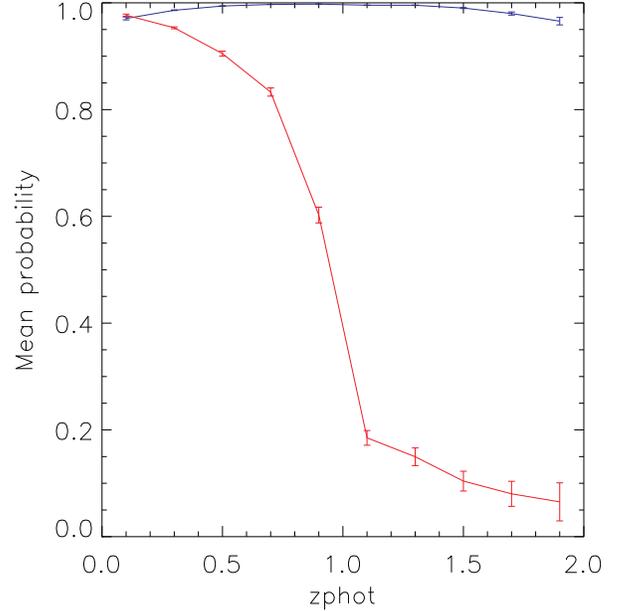}} \\
	%\mbox{\bf (a)} \\
	%\includegraphics[width=0.38\textwidth,height=0.3\textwidth]{onebyone_mag.ps} \\
        %\mbox{\bf (b)}\\
%	\includegraphics[width=0.38\textwidth,height=0.3\textwidth]{huertas_fig19.eps} &
%	\includegraphics[width=0.38\textwidth,height=0.3\textwidth]{huertas_fig18.eps} \\ 
%	\mbox{\bf (c)} & \mbox{\bf (d)}\\
%\mbox{\bf (a)} & \mbox{\bf (b)}
	%\multicolumn{1}{l}{\mbox{\bf (b)}} \\ [-0.53cm]
	
%\mbox{\bf (c)} & \mbox{\bf (d)}

%\end{array}$
\end{center}
\caption{One-to-one comparison of the ACS and WIRCam classifications as a function of redshift. Red line: mean probability for a galaxy classified as early-type in the K-band to be classified as early-type in the I-band (HST). Blue line: mean probability for a galaxy classified as late-type in the K-band to be classified as late-type in the I-band. Error bars show the dispersion of the probability distribution.}
\label{fig:onebyone}
\end{figure}

Figure~\ref{fig:onebyone} shows the mean probability for a galaxy classified as early (late) in the Ks-band to be classified as early (late) in the I-band  as a function of redshift. 

We observe, that globally, there is no ambiguity in the identification of late-type objects. A galaxy which is found to be late-type in the Ks-band is also a late-type system in the I-band with a probability around $p\sim0.9$ up to $z\sim2$. 

For early-type objects, there is clear trend with redshift: below $z\sim0.8$ a galaxy classified as early-type in the Ks-band has a probability of $p\sim0.8$ of being early-type in the I-band. Above $z\sim1$ the discrepancies between the two classifications become higher and an elliptical galaxy is classified as early in both classifications only with a probability of $p\sim0.4$. Above $z\sim1$ the I-band filter starts to probe the UV flux and therefore the obtained morphology is determined from young stars whereas the Ks-band filter stills probe the visible spectra of the galaxy. We are thus probably seeing a k-correction effect: late-type objects are identified in the same way since their stellar populations are younger whereas for early-type objects the ambiguity is higher and a fraction of objects tend to move to later morphological types. 

What are then the differences of morphologically selecting a sample in the Ks-band or in the I-band above $z\sim1$? 

\subsection{Comparing redshift distributions}

\begin{figure}[h!]
\begin{center}
%$\begin{array}{c}
%\multicolumn{1}{l}{\mbox{\bf (a)}} &
	%\multicolumn{1}{l}{\mbox{\bf (b)}} \\ [-0.53cm]
	 \resizebox{\hsize}{!}{\includegraphics{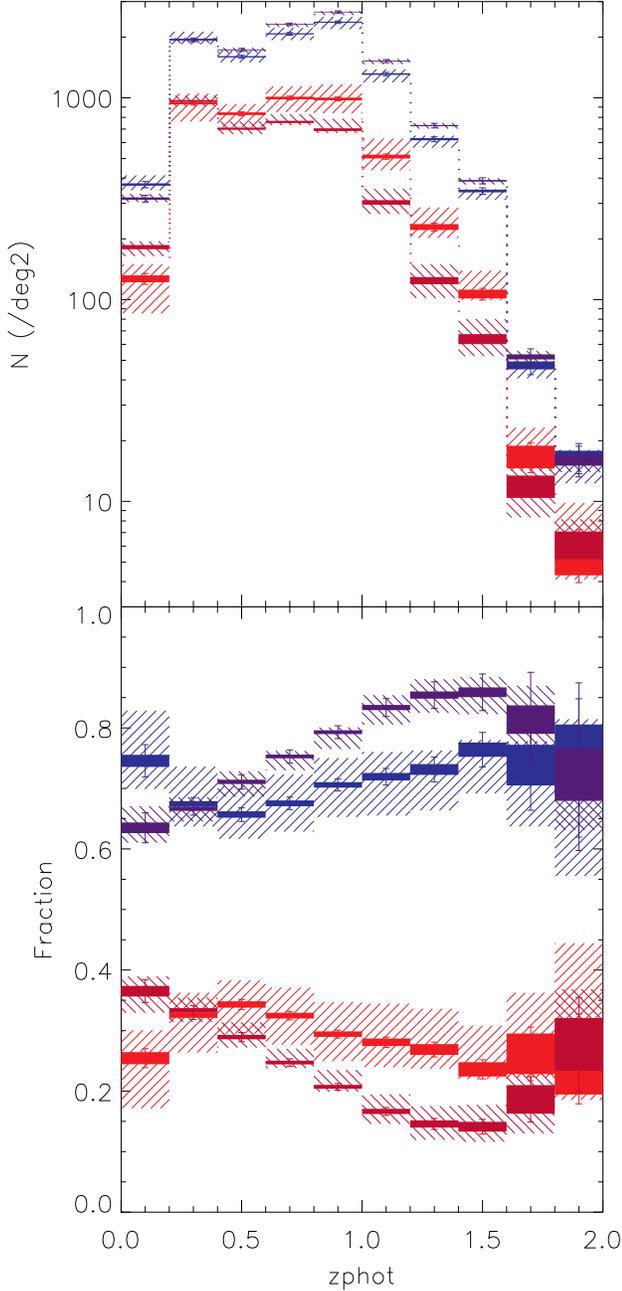}} 
	%\mbox{\bf (a)} \\
	%\includegraphics[width=0.38\textwidth,height=0.3\textwidth]{z_dist_vs_acs_34165.ps} \\ 
	%\mbox{\bf (b)}\\
%	\includegraphics[width=0.38\textwidth,height=0.3\textwidth]{huertas_fig12.eps} &
%	\includegraphics[width=0.38\textwidth,height=0.3\textwidth]{huertas_fig13.eps} \\ 
%	\mbox{\bf (c)} & \mbox{\bf (d)}\\
%\mbox{\bf (a)} & \mbox{\bf (b)}
	%\multicolumn{1}{l}{\mbox{\bf (b)}} \\ [-0.53cm]
	
%\mbox{\bf (c)} & \mbox{\bf (d)}

%\end{array}$
\end{center}
\caption {Redshift distribution per morphological type for the same $32\,171$ Ks-selected galaxies observed in the I-band and in the Ks-band. Red and pink lines are WIRCam and ACS early-type objects respectively and blue and violet lines are WIRCam and ACS late-type. The line width is the $1-\sigma$ confidence interval computed from the probability distribution, the dashed region is the confidence region deduced from Monte Carlo simulations of the training sample and the error bars are calculated using Poisson $\sqrt{n}$ statistics (see text for details).}
\label{fig:svm_comparison2}
\end{figure}

Figure~\ref{fig:svm_comparison2} shows the redshift distributions obtained from both classifications. The distributions for late-type objects present small differences as expected from the comparison performed in the previous section. Notice however that in the lowest redshift bin, the two classification find different values. This is probably a consequence of classification errors since very big galaxies are not well represented in the training. A Kolmogorov-Smirnov test (KS-test) reveals that the two functions arise from the same statistic with $96\%$ confidence. 
The early-type redshift distributions present more differences, as expected from the comparisons of the previous section, in particular above $z\sim1$. The match between the two distributions as computed from the KS-test is high at $z<1$ ($97\%$) but clearly decreases above $z\sim1$ ($67\%$). Interestingly,  the I-band estimate goes out of the confidence region at $z>1$. We find in particular an excess of early-type galaxies by a factor $\sim$1.5 in the Ks-band when compared to the I-band. Above $z\sim1.5$ the number of objects is small and the uncertainties high. Even if at $z\sim1.5$, the Ks-band classification is less accurate (\S~\ref{sec:accuracy}), the difference is significant after taking into account most of the classification errors. The differences are probably due to morphological k-correction effects. It seems therefore that a morphological classification based on HST/ACS I-band data at those redshifts tends to under-estimate the elliptical population. Nevertheless, these results have to be confirmed with a more precise study of the early-type population at $z\sim1.5$ by studying their star forming histories and mass distributions. Another possible explanation for this result is the relative sizes of the PSFs between the two
datasets: a typical galaxy in the ACS images has about fifty times more
resolution elements across it than the K-band data (since the
PSF sizes are ~0.1" vs 0.7"), so the K-band
data are better at pulling out lower surface brightness faint
envelopes, which might be missed in the ACS data because the pixels
are too small and there is too much noise. A test of this would be
to convolve the ACS data to a PSF that
matches the K-band data. We are planning this task in a forthcoming paper.

\section{Summary and conclusions}

\label{sec:summary}
\renewcommand{\labelenumi}{\roman{enumi}}
\newcommand{\tempenumi}{(\roman{enumi})}
\renewcommand{\labelenumi}{\tempenumi}
We have presented a morphological classification in two main morphological types of $44\,089$ galaxies within the COSMOS field from seeing limited near-infrared imaging. Morphologies are estimated with the non-parametric N-dimensional code galSVM using a 10 dimensional volume and non-linear boundaries. 

The final output catalog includes for every galaxy, a class label (early or late) and a probability of belonging to the class. The probability is proved to be highly correlated with the success rate and can therefore be used to assess the accuracy of the classifications. 

This classification method has been used to obtain the number counts and the redshift distribution per morphological type up to $z\sim2$ and to compare the results with the ones obtained from HST/ACS imaging on $32\,171$ galaxies in order to quantify morphological k-correction effects. \\

Our main conclusions are hereafter summarized:\\

Concerning the reliability of our morphological classification:
\begin{enumerate}

\item According to the simulated test sample, the average success rate is $\sim$80\% for the whole sample for both morphological classes, leading to $\sim$20\% contaminations ( i.e. fraction of failures).

\item The probability parameter which results from the classification procedure is a good estimator of the reliability of the classifications since it is highly correlated with the success rate.  Objects with probabilities greater than 0.8 are identified with nearly $90\%$ of confidence. 

\item The study of the probability distributions as a function of the magnitude, the size and the redshift of the galaxies reveals that the most difficult class to isolate are faint ($K_s>20$), small ($\log(r_{half})<-0.2$) early-type objects above $z\sim1$. However, even for this class of objects, the average probability is around $p\sim0.7$.

\item We also showed that selection based on a probability threshold does lead to a biased sample towards late-type systems. We proposed however a way of integrating the information brought by the probability parameter to perform statistical analysis. 

\item Errors due the training sample dominate the error source. 

\end{enumerate}

\begin{comment}
Concerning the morphological distribution of our sample:
\begin{enumerate}

\item The global morphological mixing is 30711.7$\pm$1952 ($\sim$70\%$\pm5$\%) late-type galaxies and 13376.3$\pm$2014 ($\sim$30\%$\pm5$\%) early-type galaxies. These results take into account the classification errors thanks to the use of the probability information. The galaxy population is therefore dominated by late-type systems up to $z\sim2$.

\item We detect a decrease of the fraction of early-type objects with redshift. There are $\sim32\%\pm5\%$ early-type galaxies at $z<0.4$ in our sample whereas the fraction at  $z\sim1.5$ is $21.9\%\pm8\%$. This effect persists even after correcting for classification errors using the probability parameter and therefore confirms a result obtained in previous works using HST imaging. This result probably reflects a progressive building up of the red sequence from late-type objects. However, the fraction found here at $z\sim1.5$ is significantly higher than the one obtained in previous studies \cite{Cassata05, arn07, Abraham07}. This could be explained by a selection effect since we are considering a magnitude-limited sample and will be investigated in a forthcoming paper.

\end{enumerate}

\end{comment}

We measure a global morphological mixing of 30711.7$\pm$1952 ($\sim$70\%$\pm5$\%) late-type galaxies and 13376.3$\pm$2014 ($\sim$30\%$\pm5$\%) early-type galaxies. There are $\sim32\%\pm5\%$ early-type galaxies at $z<0.4$ in our sample whereas the fraction at  $z\sim1.5$ is $21.9\%\pm8\%$. This effect persists even after correcting for classification errors using the probability parameter. The measurement
qualitatively confirms the trend observed in previous studies as a
consequence
of a progressive building up of the red sequence from late-type objects
(e.g. \citealp{Abraham07, arn07}).
We will investigate the evolution of the fraction in types defined with
our method, and using a volume limited sample rather than a magnitude
limited
sample, in a forthcoming paper.\\ 

The comparison of the morphologies with the ones obtained obtained with HST/ACS in the I-band for $32\,171$ objects reveals several interesting trends:
\begin{enumerate}

\item The global morphological mixings are globally consistent. We find $9\,163$ early-type galaxies and $25\,002$ late-type galaxies in the I-band and $10\,458$ early-type for $23\,706$ late-type galaxies in the Ks-band.

\item A galaxy classified as late-type in the I-band has a mean probability of $p\sim0.9$ of being classified as late-type in the Ks-band. 

\item The match between the two photometric bands for early-type galaxies depends on redshift: Below $z\sim1$, an early-type galaxy in the I-band has a probability of $p\sim0.7$ of being early-type in the Ks-band. Above this redshift, where the HST cameras are probing the UV flux, the probability decreases and reaches $p\sim0.4$ at $z\sim1.5$. 

\item The comparison of the redshift distributions reveals also a redshift dependence: below $z\sim1$ the two redshift distributions match quite well. The Kolmogorov-Smirnoff test gives a probability of match of $p=0.97$.
Above $z\sim1$, the I-band classification tends to find less early-type galaxies than the Ks-band one by a factor $\sim1.5$. This probably reflects a morphological k-correction effect. Therefore, studies based on HST classifications at those redshifts could underestimate the elliptical population.

\end{enumerate}

The results presented here quantify
   the bias in morphological classification due to
   morphological band shifting between I-band and K-band
   imaging data. We estimate that the fraction of missing
   early type galaxies at $z\sim1$ from I-band ACS data
   is about 19\%, increasing to 31\% at z$\sim$1.5-2.
   From our K-band data we estimate that the fraction
   of early-type galaxies has increased by 10\%
   between z=1.5 and z=0, a further confirmation
   of the gradual build up of the elliptical galaxy
   population at the expense of late-type galaxies.
   The Ks-band classification performed here is intended
   as a framework for future studies of the evolution of
   counts, luminosities, luminosity densities and correlation
   function for each morphological type over several
   redshift bins on a volume limited sample.

\begin{acknowledgements}
The authors want to thank the anonymous referee for his useful suggestions that clearly improved the paper. This work is part of a big collaboration, we would like to thank as well all the people who contributed somehow to this paper, specially with the data reduction (H. Aussel, D. Thompson, O. Ilbert).  
\end{acknowledgements}

\bibliographystyle{aa}
\bibliography{1255biblio.bib}

\end{document}